\documentclass[amsmath,amssymb,aps,prx,reprint,superscriptaddress,nofootinbib]{revtex4-2}
\usepackage{graphicx}
\usepackage[caption=false]{subfig}
\usepackage{color}

\newtheorem{theorem}{Theorem}
\newtheorem{assumption}[theorem]{Assumption}

\newtheorem{corollary}[theorem]{Corollary}

\newtheorem{example}{Example}
\newtheorem{lemma}[theorem]{Lemma}

\newcommand{\df}[2]{\frac{d \, #1}{d #2}}

\newcommand{\pd}[2]{\dfrac{\partial #1}{\partial #2}}
\renewcommand{\P}{\textnormal{Pr}}

\renewcommand{\O}{\mathcal{O}}

\newcommand{\eps}{\varepsilon}

\newcommand{\s}{\mathbf{s}}
\renewcommand{\v}{\mathbf{v}}
\newcommand{\w}{\mathbf{w}}
\newcommand{\x}{\mathbf{x}}
\newcommand{\y}{\mathbf{y}}
\newcommand{\ER}{Erd\H{o}s-R\'{e}nyi}
\newcommand{\LTset}{\Theta} % \theta, \xi, \varphi, \psi, \omega
\newcommand{\LTel}{\theta}
\newcommand{\NTset}{\Xi} % \theta, \xi, \varphi, \psi, \omega
\newcommand{\NTel}{\xi}

\newcommand{\quotes}[1]{``#1''}

\usepackage{tikz}
\usetikzlibrary{arrows,automata,fit,shapes,arrows.meta,positioning}

\begin{document}
	
%\preprint{}

%\title{Exact solution of heterogeneous Markovian SI epidemics on networks}
%\title{Exact analytic solution of Markovian compartmental epidemics without cycles on networks}
\title{Analytic solution of Markovian epidemics without re-infections \\ on heterogeneous networks}


\author{Massimo A. Achterberg}
\email{M.A.Achterberg@tudelft.nl}
\affiliation{Faculty of Electrical Engineering, Mathematics and Computer Science, Delft University of Technology, P.O. Box 5031, 2600 GA Delft, The Netherlands}
\author{Piet Van Mieghem}
\affiliation{Faculty of Electrical Engineering, Mathematics and Computer Science, Delft University of Technology, P.O. Box 5031, 2600 GA Delft, The Netherlands}
%\thanks{}


\date{\today}

\begin{abstract}
	%We study heterogeneous, continuous-time Markovian Susceptible-Infected (SI) epidemics on a network with $N$ nodes, which is a pure infection process without curings. Each node is either susceptible or infected at any time, leading to $2^N$ possible configurations. The corresponding infinitesimal generator $Q$ is a $2^N\times 2^N$ upper triangular matrix, whose diagonal elements are equal to its eigenvalues. We show that the eigenvalues of the infinitesimal generator $Q$ are equal to the sum of all weighted links in the cut set between all susceptible and infected nodes. If the eigenvalue of a certain configuration is small (in absolute value), it takes a long time to spread the epidemic to an adjacent node. Additionally, we exactly compute the eigenvectors of the infinitesimal generator~$Q$ in an iterative way. Assuming that all eigenvalues of the infinitesimal generator $Q$ are distinct, the time-dependent solution of the Markov chain is computed exactly. This contrasts the mean-field approximation of the SI process, which cannot be solved analytically. We verify our exact formula by simulations on several small graphs. The versatility of our exact method is demonstrated by generalising the SI process to non-Markovian dynamics, temporal networks, simplicial contagion and self-infections. Finally, we show that the same procedure can be applied to SIR dynamics.
	
	Most epidemic processes on networks can be modelled by a compartmental model, that specifies the spread of a disease in a population. The corresponding compartmental graph describes how the viral state of the nodes (individuals) changes from one compartment to another. If the compartmental graph does not contain directed cycles (e.g.\ the famous SIR model satisfies this property), then we provide an analytic, closed-form solution of the continuous-time Markovian compartmental model on heterogeneous networks. The eigenvalues of the Markovian process are related to cut sets in the contact graph between nodes with different viral states. We illustrate our finding by analytically solving the continuous-time Markovian SI and SIR processes on heterogeneous networks. We show that analytic extensions to e.g.\ non-Markovian dynamics, temporal networks, simplicial contagion and more advanced compartmental models are possible. Our exact and explicit formula contains sums over all paths between two states in the SIR Markov graph, which prevents the computation of the exact solution for arbitrary large graphs.
\end{abstract}

%\keywords{}

\maketitle


% If in two-column mode, this environment will change to single-column
% format so that long equations can be displayed. Use
%\begin{widetext}
%\end{widetext}

%\section*{Popular summary}
%Modelling the spread of infectious diseases received a lot of attention during the COVID-19 pandemic. Basic compartmental models of infectious diseases are the Susceptible-Infected (SI), the Susceptible-Infected-Susceptible (SIS) and the Susceptible-Infected-Recovered (SIR) model, which categorise individuals in the compartments (S) susceptible, but healthy, (I) infected and (R) recovered. We investigate the spread of an infectious disease in a heterogeneous population. Despite the simplicity of the SIS and SIR Markovian models, exact solutions on heterogeneous contact networks have not been found so far. 

%For the first time, we present here an exact, analytic solution of the SI and SIR models in a heterogeneous population. Using the exact solution, we precisely determine the time at which the maximum number of infected cases is attained in SIR epidemics. We believe that the exact solution paves the way for solving more advanced and realistic epidemic models with a more complex disease properties. If quantum computers and quantum algorithms are effective, then our exact, analytic solution is computable for real-world contact graphs.

\section{Introduction}
Most research in network epidemiology focuses on so-called compartmental models, in which the population is subdivided into several groups based on the current viral state of the individual \cite{ReviewPaperSIS}. The SIS and SIR models are the most elementary compartmental models, in which (S) represents a susceptible, but healthy individual, (I) represents an infected and infectious individual and (R) represents a resistant or removed individual, who is no longer able to contract the disease. Both SIS and SIR are composed of two processes: (i) infected individuals can infect susceptible neighbours and (ii) infected individuals can recover, either becoming susceptible again (SIS) or receiving permanent immunity (SIR).

% What is a network + notation
The individuals can be represented by a set $\mathcal{N}$ of $N$ nodes in a contact graph $G$. The set $\mathcal{L}$ of $L$ links describes the connections between nodes with a certain weight. The elements $a_{ij}$ of the $N \times N$ adjacency matrix~$A$ denote whether there is a link ($a_{ij}=1$) between nodes $i$ and $j$ or not ($a_{ij}=0$). We assume that the contact graph~$G$ is connected, such that any pair of nodes is connected by a path in the contact graph. 

% General compartmental models
General epidemic models are comprised of $c$ compartments. We assume that interactions only occur between neighbouring nodes and that after such an interaction, at most one of those nodes changes its viral state. Then the transitions between compartments can be classified into two types \cite{GEMF2}: link-based transitions, such as the infection of a susceptible node by a neighbouring infected node and node-based transitions, such as nodes recovering from the disease. Many results in epidemics have been obtained in the well-mixed limit, i.e.\ all individuals can directly contact all other individuals \cite{anderson_may_1992}. In reality, each individual has different characteristics and cannot contact any other member of the population, leading to heterogeneous behaviour during an epidemic \cite{ReviewPaperSIS}. Therefore, we focus in this work on epidemic processes on heterogeneous networks.

% What others do: Monte Carlo and mean-field
Properties of such network-based models are usually computed by mean-field approximations \cite{SIScomputervirus} or Monte Carlo simulations \cite{cota2017simulatingmarkov}. The downside is that Monte Carlo simulations are known to converge slowly, which is problematic if the probability of a certain event is small. Simulating is especially tricky around the epidemic threshold, which is precisely the regime in which most real-world epidemics occur. On the other hand, the mean-field approximation works well for homogeneous, well-mixed contact graphs but is less precise for heterogeneous networks \cite{AccuracySIS}. Instead of resorting to either Monte Carlo simulations or mean-field approximations, we focus in this work on determining the \emph{exact} time-dependent solution for general Markovian compartmental models.

% Literature overview on infinitesimal generators
For the computation the exact time-dependent solution, we require an efficient enumeration of all states of the underlying Markov chain. Since each of the $N$ nodes can attain $c$ viral states, the number of states equals $c^N$. The set of all possible states is denoted by $\mathbf{S}$. Providing explicit constructions for the states\footnote{Throughout this work, we will use \emph{state} and \emph{configuration} when referring to one element of the state space $\mathbf{S}$, interchangeably.} and the corresponding infinitesimal generator on heterogeneous networks for general Markovian compartmental models is a tedious task. Several researchers have attempted to construct infinitesimal generators for specific epidemic models on networks. For example, Van Mieghem \emph{et al}.\ \cite{VanMieghemOmicKooij2009} provided a construction based on binary numberings for SIS epidemics and for SIS epidemics with self-infections \cite{IntroductionEpsSIS}. Simon \emph{et al}.\ \cite{Simon2011} suggested a tridiagonal block structure for the infinitesimal generator of the SIS process. Each block contains configurations with the same number of infected nodes $k$. Configurations with $k$ infected nodes can only make transitions to configurations with $k+1$ and $k-1$ infected nodes, such that the infinitesimal generator $Q$ has a block-tridiagonal structure. Economou \emph{et al}.\ \cite{ExactSISconstruction} developed a similar tridiagonal block structure, but used a different ordering within each block. For SIR epidemics, L\'{o}pez-Garc\'{i}a \cite{ExactSISconstruction2} groups configurations based on the number of recovered nodes. Within each block with the same number of recovered nodes, configurations are grouped based on the number of infected nodes. Within each block with the same number of recovered \emph{and} infected nodes, a lexicographical ordering is applied. The Generalised Epidemic Mean-Field (GEMF) model by Sahneh \emph{et al}.\ \cite{GEMF2} describes disease transmission in general compartmental models, where the state space is constructed using a tensor or Kronecker product formulation. Merbis \cite{merbis2021exact} used with a similar tensor product construction to derive the time-dependent equations for general epidemic models. In this tensor formulation, Merbis and Lodato \cite{merbis2021logistic} derived an exact solution for the SI process on unweighted graphs. Another representation in terms of a tensor-product formulation for SIR epidemics is provided by Dolgov and Savostyanov \cite{dolgov2022SIR3N}.

% Relation to event graphs in temporal networks
Mappings of a temporal network to a directed, static graph have been established. An example of such a directed, static graph is the event graph \cite{badie2022directedtemporal}, whose nodes represent events (e.g.\ corresponding to the creation or removal of links in the contact graph) and its links represent possible follow-up events. The space of all states in Markovian epidemics has a similar structure, although nodes and links have a different meaning. We believe that our solution method is superior, because it can also include temporal networks, whereas an extension of the event graph to handle epidemic spreading is not straightforward.

% What we do and why
General compartmental epidemic models, as formalised in the GEMF framework \cite{GEMF2}, can be solved by the eigendecomposition of the corresponding infinitesimal generator. If the compartmental graph is a tree, i.e.\ the compartmental graph does not contain any cycles, then the corresponding Markov graph is also a tree, whose infinitesimal generator $Q$ only has upper-triangular non-zero elements. Consequently, the eigenvalues and eigenvectors can be determined efficiently. If cycles occur, then the infinitesimal generator $Q$ is no longer upper-triangular and our elegant solution method is not applicable, because the eigenvalues and eigenvectors can only be determined numerically, which is infeasible for large graphs.

% Structure
In this work, we present for the first time an analytic closed-form solution of  transient continuous-time Markovian epidemic processes, i.e. without re-infections, on heterogeneous networks. We demonstrate our method by considering SI epidemics in Section~\ref{sec_SI_process}. We present an in-depth explanation of the eigenvalues of the infinitesimal generator in Section~\ref{sec_eigenvalues} and derive the full solution in Section~\ref{sec_exact_solution}. We discuss the benefits and difficulties of the exact solution in Section \ref{sec_comp_infeasible} and compare our exact results with simulations in Section \ref{sec_simulations}. We then generalise our exact solution on non-Markovian dynamics, temporal networks, simplicial contagion and self-infections in Section~\ref{sec_extensions}. We also provide an extensive analysis on SIR epidemics in Section~\ref{sec_SIR}: the SIR eigenvalues in Section~\ref{sec_SIR_eigenvalues}, SIR eigenvectors in Section~\ref{sec_SIR_eigenvectors} and the full solution in Section~\ref{sec_SIR_solution}. We demonstrate the power of our exact method by analytically computing several properties in Section~\ref{sec_SIR_peak}-I. Finally, we conclude in Section \ref{sec_conclusion}.

\section{The Markovian SI process}\label{sec_SI_process}
The Susceptible-Infected (SI) process describes the spread of information, diseases, innovations or neural activity over a network. Starting from a single infected node, the infection spreads to all its neighbours, which again infects their neighbours, until the whole network is infected \cite{BaileySI}. The Markovian SI process is actually a Markovian discovery process, or, equivalently, a stochastic shortest path process \cite[Chapter 16]{PA}. The SI process describes spreading phenomena without damping or curing (for example, HIV viruses in humans or the spread of innovations) or the spread evolves extremely fast, effectively flooding the population and  prohibiting curing or recovering (for example, epileptic seizures in the human brain \cite{stam2022seizure} and the spread of information on social media).

\begin{figure*}[!htb]
    \centering
    \subfloat[\label{fig_Markovgraph1} Compartmental graph]{%
        \includegraphics[width=0.33\textwidth]{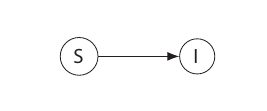}
    }
    \subfloat[\label{fig_Markovgraph2} Contact graph]{%
        \includegraphics[width=0.21\textwidth]{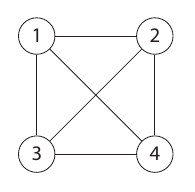}
    }
    \subfloat[\label{fig_Markovgraph3} Markov graph]{%
        \includegraphics[width=0.44\textwidth]{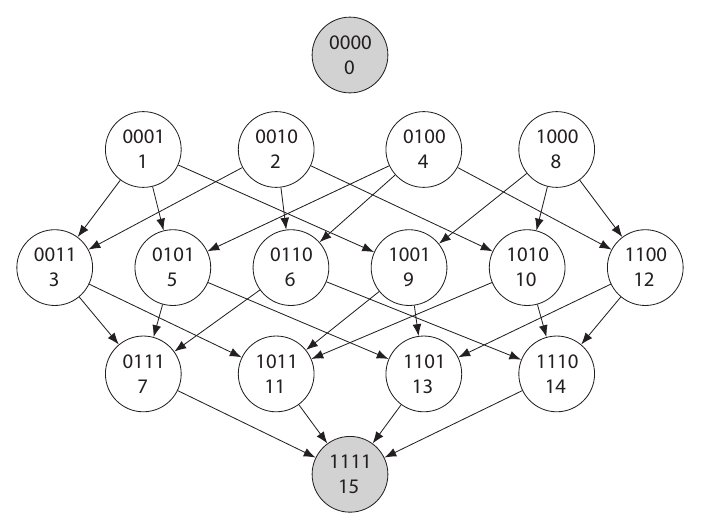}
    }
    \caption{The spread of SI epidemics involves three networks: (a) the compartmental graph specifies the transitions between the different compartments, (b) the nodes in the contact graph $G$ represent individuals and the links specify the connections among the individuals and (c) the combination of the SI process from (a) and the contact network from (b) leads to the Markov graph containing all possible configurations and the transitions between those configurations. Grey nodes indicate absorbing states. The number of compartments is 2 and the number of nodes $N$, such that the Markov graph is of size $2^N$. In this example, $N=4$.}
    \label{fig_Markovgraph_SI}
\end{figure*}

The SI process subdivides the nodes into two compartments: the set $\mathcal{I} \subseteq \mathcal{N}$ of infected (I) nodes are infected, aware or activated and the complementary set ${\cal S} = {\cal N}\backslash {\cal I}$ of susceptible (S) nodes are healthy, unaware or idle. We further assume that an infection over the direct link from node $i$ towards a susceptible node~$j$ is a Poisson process with rate $\beta_{ij}$ and is independent of all other links. The associated compartmental graph is shown in Figure~\ref{fig_Markovgraph1}. The SI process is a continuous-time Markov process containing $2^N$ states, because each of the $N$ nodes in the contact graph is either susceptible or infected. The $2^N \times 1$ state vector is $\s(t) = (s_0(t), s_1(t), \ldots, s_{2^N-1}(t))^T$, whose element $s_i(t)$ describes the probability that the SI process is in state~$i$ at time $t$. Since the process must certainly be, for any time $t$, in one of the possible states, $\sum_{i=0}^{2^N-1} s_i(t) = 1$ for all times $t$. The transition rates between the states are denoted by the $2^N \times 2^N$ infinitesimal matrix $Q$.

We describe the SI process using the labeling in \cite{VanMieghemOmicKooij2009}. The idea of the construction in \cite{VanMieghemOmicKooij2009,IntroductionEpsSIS} is that the state $i$ represents the viral state of $N$ nodes in the contact graph. We define the binary variables~$x_k(i)$ indicating whether node~$k$ is infected in configuration $i$. Then the \emph{viral state vector} $\mathbf{x} = (x_N, x_{N-1}, \ldots, x_2, x_1)^T$ describes the viral state of all nodes. By regarding the viral state vector $\mathbf{x}$ as a binary number, the configuration number $i$ is computed as
\begin{equation*}
	i = \sum_{k=1}^N x_k(i) 2^{k-1},
\end{equation*}
where $x_k=1$ if node $k$ is infected and $x_k=0$ if node $k$ is susceptible. For a complete contact graph with $N=4$ nodes, the possible transitions for the SIS process are shown in Figure \ref{fig_Markovgraph_SI}. For example, the all-healthy state $i=0$ has representation $(0000)$, indicating that all nodes in state $0$ are healthy. Similarly, for state $5$ the binary representation is $(0101)$, indicating that node 2 and 4 are healthy and node 1 and 3 are infected.

A fundamental property of Markov processes is that events in continuous time occur sequentially and simultaneous events do not happen almost surely. This means that transitions between states in the Markov chain can only occur if the binary representation of those states differs exactly one bit, which is illustrated in Figure \ref{fig_Markovgraph3}. However, not all states differing one bit are connected in Figure~\ref{fig_Markovgraph3}. The reason is that the SI process only considers an infection process: upward-pointing transitions in the Markov graph in Figure~\ref{fig_Markovgraph3} cannot occur, because upward transitions correspond to the recovery of a node. Furthermore, if the SI process is in the all-healthy state, then the spread never activates, because there is no initial infection. Hence, the all-healthy state $i=0$ cannot be entered nor left, thus state $i=0$ has no incoming nor outgoing arrows in the SI process. Finally, we remark that the contact graph $G$ in Figure~\ref{fig_Markovgraph2} describes the existence of links between pairs of nodes. Naturally, the existence and weight of the links in the Markov graph are influenced by the existence of links in the contact graph $G$. 

\begin{figure}[!ht]
	\centering
	\includegraphics[width=0.8\columnwidth]{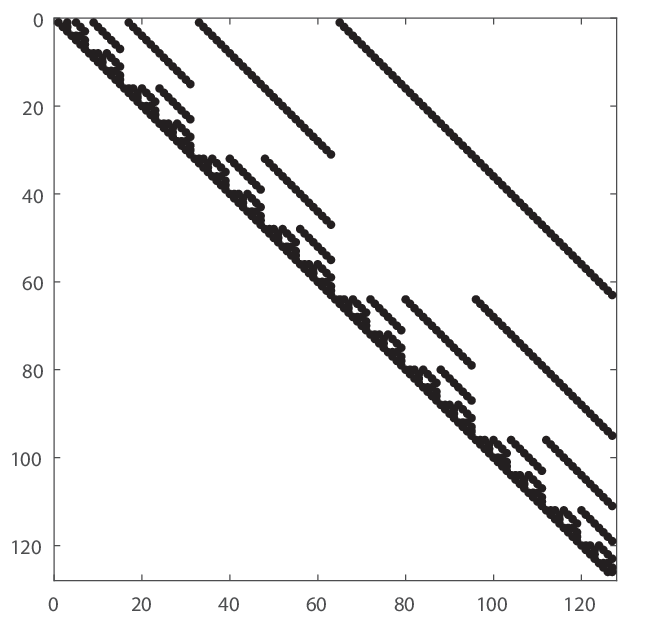}
	\caption{The structure of the infinitesimal generator $Q$ for the SI process on a complete graph with $N=7$ nodes.}
	\label{fig_Q}
\end{figure}

Using the binary notation of the state space $\mathbf{S}$, we describe the transitions between the states using the $2^N \times 2^N$-dimensional infinitesimal generator $Q$. To simplify the notation, we introduce the $N \times N$, possibly asymmetric, weighted and non-negative adjacency matrix $B$ with elements $\tilde{\beta}_{kl} = a_{kl} \beta_{kl}$. The elements $q_{ij}$ of the infinitesimal generator $Q$ for the SI process are \cite[Chapter 17]{PA}
\begin{subequations}
	\begin{align}
		q_{ij} &= \sum_{k=1}^N \tilde{\beta}_{mk} x_k(i), \qquad \text{if } j = i + 2^{m-1} \\
		&\qquad\qquad \text{ with } m = 1,2,\ldots, N \text{ and } x_m(i)=0. \nonumber \\
		q_{ii} &= - \sum_{\substack{j=0 \\ j\neq i}}^{2^N-1} q_{ij},
	\end{align}
\end{subequations}
for $i,j = 1, 2, \ldots, 2^N$. The non-zero elements of the infinitesimal generator $Q$ are visualised in Figure~\ref{fig_Q}. The infinitesimal generator $Q$ is an upper triangular matrix, because in absence of curing events, there is no transition from state $i$ to state $j<i$ because the number of infected nodes is larger for state $i$ than for state $j$. The infinitesimal generator $Q$ for a contact graph with $N$ nodes can also be constructed recursively in terms of the infinitesimal generator of a graph with $N-1$ nodes and then adding one node \cite{IntroductionEpsSIS}.
The governing equation for the continuous-time Markov chain is
\begin{equation}\label{eq_MC_diff}
	\df{\s^T(t)}{t} = \s^T(t) Q,
\end{equation}
whose solution is
\begin{equation*}
	\s^T(t) = \s^T(0) e^{Q t}.
\end{equation*}
The solution can be further detailed as \cite[p. 4]{VanMieghemOmicKooij2009}
\begin{equation}\label{eq_MC_general_solution}
	\s(t) = \pi + \sum_{i=1}^{2^N-1} e^{\lambda_i t} \sum_{m=0}^{n_i-1} \mathbf{r}_{i,m} \frac{t^m}{m!}.
\end{equation}
where $n_i$ denotes the multiplicity of eigenvalue $\lambda_i$, $\pi = (0, 0, \ldots, 0, 1)^T$ is the steady-state vector and the vector $\mathbf{r}_{i,m}$ is related to the right- and left-eigenvectors of $Q$ and the initial condition $\s(0)$. Unfortunately, the infinitesimal generator $Q$ is asymmetric and is not even normal\footnote{An $N\times N$ matrix $A$ is \emph{normal} if it commutes with its conjugate transpose: $A A^* = A^* A$.}, preventing further simplifications in Eq.\ (\ref{eq_MC_general_solution}).

The SI process is completely described by the set (\ref{eq_MC_diff}) of $2^N$ linear differential equations. Frequently, an exponentially large state space is the fingerprint of a non-linear process. In particular, SIS and SIR epidemics exhibit a phase transition around the epidemic threshold. Below the threshold, the epidemic dies out exponentially fast, whereas above threshold, a non-zero fraction of the population remains infected over a long time. On the contrary, the SI process does not exhibit a phase transition, because the SI process always converges -- given that the process starts with at least one infected node --  to the all-infected state, irrespective how small, but non-zero, the infection rates $\beta_{ij}>0$ are. For simple graphs with homogeneous transition rates\footnote{Homogeneous infection rates are $\beta_{kl} = \beta$ for all nodes $k,l$, such that $B = \beta A$.}, like the complete graph or the star graph, an exact analysis is possible \cite{DefinitionMetastableState} by exploiting symmetry of the contact graph \cite{Simon2011}. Otherwise, exactly solving equation (\ref{eq_MC_diff}) or (\ref{eq_MC_general_solution}) for the SI process is infeasible for large $N>20$ and one resorts to Monte Carlo simulations \cite{MonteCarloSIS,cota2017simulatingmarkov}.

The general solution (\ref{eq_MC_general_solution}) contains the eigenvalues and eigenvectors of the infinitesimal generator~$Q$. The eigenvalues $\lambda_i$ of the infinitesimal generator $Q$ are of primary importance in the solution (\ref{eq_MC_general_solution}), because the eigenvalues are inversely proportional to the relaxation time of the corresponding eigenmode. The eigenvalue $\lambda_i$ is therefore related to the average time for the SI process to make a transition from configuration $i$ to another configuration~$j$. The next section further elaborates on the eigenvalues of the SI process.

\subsection{Eigenvalues of the infinitesimal generator $Q$}
\label{sec_eigenvalues}
The $2^N$ eigenvalues of the $2^N \times 2^N$ infinitesimal generator $Q$ of a Markov chain have negative real part and at least one eigenvalue is zero \cite{PA}. Since the infinitesimal generator of the SI process is upper triangular as illustrated in Figure \ref{fig_Q}, the eigenvalues of the SI process appear on the diagonal of the infinitesimal generator $Q$, hence, $\lambda_i = q_{ii} = -\sum_{j=i+1}^{2^N} q_{ij}$ for all $i=0,1,\ldots,2^N-1$. Due to the triangular structure of the infinitesimal generator~$Q$, the eigenvalue~$i$ can be understood as to belong to configuration~$i$. Prior to describing the physical significance of the eigenvalue $\lambda_i$, we introduce some notation.

Each configuration $i$ describes which nodes in the network are infected and which nodes are susceptible. For a given configuration $i$, we partition the set of nodes $\mathcal{N}$ into two groups. The group $\mathcal{I}_i \subseteq{\cal{N}}$ defines the set of infected nodes in configuration $i$. Similarly, $\mathcal{S}_i$ contains the susceptible nodes in configuration $i$. Each node is either infected or susceptible, so $\mathcal{N} = \mathcal{I}_i \cup \mathcal{S}_i$ for all configurations $i$. We define the \emph{cut set} as the set of links with one node in $\mathcal{S}_i$ and one node in $\mathcal{I}_i$. 

\begin{figure}[!ht]
    \centering
    \includegraphics[width=0.8\columnwidth]{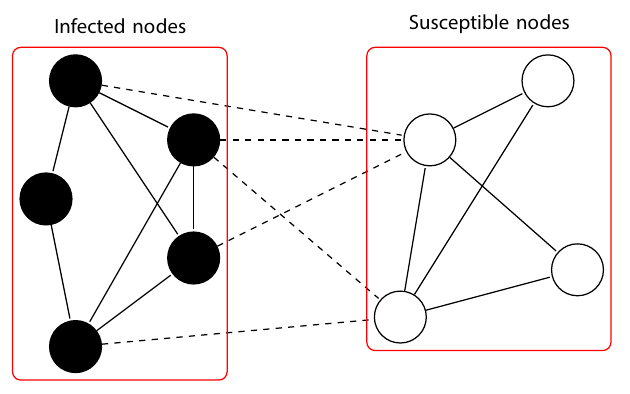}
    \caption{The cut set in SI epidemics on a contact graph with 9 nodes. The eigenvalue $\lambda_i$ that belongs to configuration $i$ is (minus) the sum over all weighted links in the S-I cut set. The links in the cut set are illustrated by dashed lines.}
    \label{fig_cut_set}
\end{figure}

The eigenvalue $\lambda_i$ is equal to the sum over all possible transitions from configuration $i$ to any other configuration $j$, where configuration $j$ differs exactly one bit (one node) from configuration~$i$. Starting in configuration $i$, visualised in Figure \ref{fig_cut_set}, the possible configurations~$j$ are those where one susceptible node will be infected by one of the already infected nodes. Hence, the eigenvalue $\lambda_i$ equals (minus) the sum over all weighted links in the S-I cut set;
\begin{equation}\label{eq_eigenvalues1}
	\lambda_i = -\sum_{k \in \mathcal{I}_i} \sum_{l \in \mathcal{S}_i} \tilde{\beta}_{kl},
\end{equation}
or, following the notation of \cite{PA};
\begin{equation}\label{eq_eigenvalues2}
	\lambda_i = -\sum_{k=1}^N \sum_{l=1}^N x_k(i) (1-x_l(i)) \tilde{\beta}_{kl}.
\end{equation}
If the contact graph $G$ has homogeneous link weights such that $\tilde{\beta}_{kl} = \beta a_{kl}$, where $a_{kl}$ specifies whether a link exists between node $k$ and $l$, then $\lambda_i / \beta$ \emph{equals (minus) the number of links in the S-I cut set}.

Equation (\ref{eq_eigenvalues1}) provides the exact eigenvalue $\lambda_i$ for every configuration $i$. The time-dependent solution (\ref{eq_MC_general_solution}) shows the implication of the eigenvalues $\lambda_i$. If the eigenvalue is large (in modulus), the corresponding state will converge exponentially fast to zero. For small (in modulus) eigenvalues, the convergence is much slower. The SI process can be regarded as a discovery process, that starts with a set of discovered nodes and discovers adjacent nodes. If $t \to\infty$ on a connected graph, all nodes will be discovered. Large (in modulus) eigenvalues have a small contribution to the total discovery time, but small (in modulus) eigenvalues will have a significantly larger contribution to the total discovery time in the SI process. In particular, a small (in modulus) eigenvalue $\lambda_i$ of a configuration $i$ corresponds to a small weight of the S-I cut set (see Figure \ref{fig_cut_set}) and it takes a long time to transfer the disease between these two groups of nodes. The minimal eigenvalue (equivalent to finding the minimum weighted cut in the contact graph) can be obtained efficiently using the Stoer-Wagner algorithm \cite{stoerwagneralgorithm}.

Finally, equation (\ref{eq_eigenvalues1}) illustrates that eigenvalue $\lambda=0$ occurs twice; the all-healthy state (where $\mathcal{I}_i = \emptyset$) and the all-infected state (where $\mathcal{S}_i = \emptyset$) have an empty cut set and the corresponding eigenvalue is zero. Hence, the all-healthy state and the all-infected are both steady states of the SI process. The all-healthy state is unstable, because adding a single infected node will lead to more infected nodes, whereas the all-infected state is stable.

Given a configuration $i$, then the number of infected nodes $k$ in configuration $i$ can be calculated as
\begin{equation}
	k = \sum_{l=1}^N x_l(i).
\end{equation}
For several contact graphs, the eigenvalues can be computed analytically. 

For the \textbf{complete graph} with homogeneous infection rates, the eigenvalues are
\begin{equation*}
	\lambda = -\beta k (N-k),\qquad \text{with multiplicity } \binom{N}{k}
\end{equation*}
where $0 \leq k \leq N$ is the number of infected nodes in configuration $i$. The SI process on the complete graph with homogeneous infection rates and starting with a fixed number $k$ of infected nodes is presumably the easiest SI process, because in this case, the SI process can be transformed to a birth and death process, whose time-dependent solution can be calculated exactly \cite[Ch.\ 16]{PA}. Starting from $k$ infected nodes and ending with $l$ infected nodes, the total spreading time is distributed as the sum of independent exponential random variables, with parameters $\lambda_m$ where $k \leq m \leq l-1$ represents the number of infected nodes \cite{BaileySI,SIexponentialvars}.

For the \textbf{star graph} with homogeneous infection rates, the eigenvalues with multiplicity $\binom{N-1}{k}$ are
\begin{equation*}
	\lambda = \begin{cases}
		-\beta (N-1-k),\qquad &\text{if hub infected} \\
		-\beta k,\qquad &\text{if hub healthy}
	\end{cases}
\end{equation*}
where $0\leq k \leq N-1$ is the number of infected leaf nodes in configuration $i$. In this case, the eigenvalues can be simplified to the form $\lambda = - \beta k$ with multiplicity $2 \binom{N-1}{k}$ for $0 \leq k \leq N-1$.

For the \textbf{cycle graph} with homogeneous infection rates, the eigenvalues are
\begin{equation*}
	\lambda_i = - 2 \beta p(k,i),
\end{equation*}
where $p(k,i)$ depends on the number of susceptible neighbours of the $k$ infected nodes in configuration~$i$. Contrary to the complete graph and the star graph, the position of the infected nodes is crucial in the cycle graph. Starting in configuration $i$ with $k$ infected nodes, the $k$ infected nodes can be grouped into $p \leq k$ connected, infectious components. The total number of susceptible neighbours is $2p$, because each infected component connects to two susceptible neighbours. After infecting one of the susceptible neighbours, the number of connected, infected components either stays constant or reduces by one, because two components will merge if the in-between susceptible node has been infected. In that case, the number of components reduces to $p-1$. The number of connected, infected components can therefore only decrease over time. Since the eigenvalue $\lambda_i$ only decreases (in modulus) over time, the discovery speed of the SI process also slows down over time.

\begin{figure}[!ht]
	\centering
	\includegraphics[width=0.98\columnwidth]{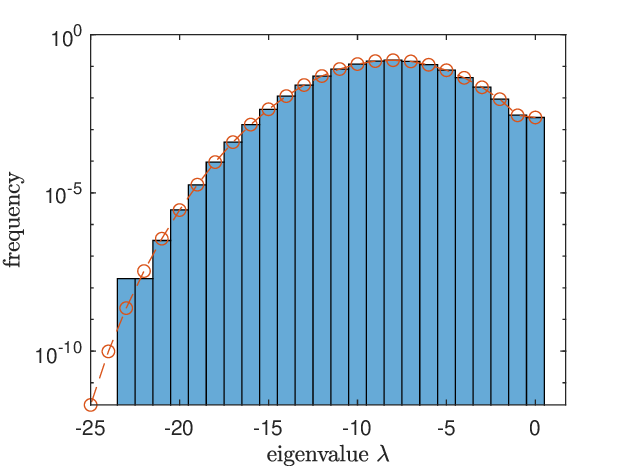}
	\caption{The eigenvalue distribution of the infinitesimal generator $Q$ of the SI process in an \ER{} graph with $N=10$ nodes and link-existence probability $p=0.36$ averaged over 100,000 simulations. The exact solution (\ref{eq_eig_ER}) is shown as a red dashed line. The simulation results do not follow the exact solution for $\lambda \leq -20$ because the number of simulations is too small compared to the probability of occurrence of these eigenvalues.}
	\label{fig_ER_eig}
\end{figure}

In the \textbf{\ER{} graph} \cite{ErdosRenyiRandomGraphs} with homogeneous infection rates with weight $\beta=1$, links exist independent of other links with probability $p$. The eigenvalue probability distribution $\P[\lambda=-l]$ of a random configuration~$i$ follows from the law of total probability;
\begin{align*}
	\P[\lambda = -l] &= \sum_{k=0}^N \P\left[\lambda = -l \ | \ \text{state } i \text{ has } k \text{ infected nodes}\right] \\
	&\qquad\quad\times \P[ \text{state } i \text{ has } k \text{ infected nodes}].
\end{align*}
Suppose that configuration $i$ consists of $k$ infected nodes. Then the eigenvalue $\lambda_i = -l$ is (minus) the number of links in the cut set of configuration~$i$. The cut set contains maximally $k(N-k)$ links, where each link exists independently of the other links with probability $p$. Hence, the probability distribution is a Binomial distribution with $k(N-k)$ possible links and probability $p$;
\begin{align*}
	&\P[\lambda = -l \ | \ \text{state } i \text{ has } k \text{ infected nodes}] \\
	&\quad=\binom{k(N-k)}{l} p^{l} (1-p)^{k(N-k)-l}.
\end{align*}
The probability for configuration $i$ to have $k$ infected nodes is equal to the probability for any graph with $N$ nodes to have $k$ infected nodes, which equals $2^{-N} \binom{N}{k}$. Combining all, we find
\begin{align}\label{eq_eig_ER}
	&\P[\lambda = -l] = \nonumber \\
	&\left(\frac{p}{1-p}\right)^l 2^{-N} \sum_{k=0}^N \binom{k(N-k)}{l} \binom{N}{k} (1-p)^{k(N-k)},
\end{align}
with the convention that $\binom{k(N-k)}{l}$ is zero if $l < 0$ or $l > k(N-k)$. Hence, the eigenvalue~$\lambda$ is bounded between $\lambda=0$ and $\lambda=-k(N-k)$. We plot the exact solution (\ref{eq_eig_ER}) with numerical simulations in Figure \ref{fig_ER_eig}, which shows excellent agreement.

\subsection{The exact SI solution}
\label{sec_exact_solution}
An explicit form of the SI solution (\ref{eq_MC_general_solution}) requires the computation of the eigenvectors of the infinitesimal generator $Q$. We denote by $\v_0, \v_1, \ldots, \v_{2^N-1}$ and $\w_0, \w_1, \ldots, \w_{2^N-1}$ the right- and left-eigenvectors of $Q$, respectively. The left-eigenvector $\w_{2^N-1}$, corresponding to the all-infected state $i=2^N-1$ with eigenvalue $\lambda_{2^N-1}=0$, is equal to the steady-state vector $\pi$:
\begin{equation}\label{eq_steady_state}
	\w_{2^N-1} = \pi = (0,0,\ldots, 0, 1)^T.
\end{equation}
The corresponding right-eigenvector is the all-ones vector $\v_{2^N-1}=(1,1,\ldots, 1)^T$. Since there are no transitions possible to and from the all-healthy state $i=0$ in the SI process, \emph{we remove the all-healthy state from the Markov chain}, which reduces the Markov chain to $2^N-1$ states.

The SI process for homogeneous transitions rates was recently solved by Merbis and Lodato \cite{merbis2021logistic}, thus we focus on heterogeneous infection rates $\tilde{\beta}_{kl}$. In many practical applications, the weighted infection rates $\tilde{\beta}_{kl}$ are real numbers and we can safely assume the following:

\begin{assumption}\label{ass_eig_distinct}
	The eigenvalues of the infinitesimal generator $Q$ are distinct.\footnote{The eigenvalues may also be degenerate, as long as the algebraic multiplicity equals the geometric multiplicity for all eigenvalues. Then the matrix is diagonalisable and Eq.\ (\ref{eq_general_solution}) holds. It is, however, unclear when the algebraic and geometric multiplicities are equal for general contact graphs.}\textsuperscript{,}\footnote{Suppose that the weighted infection rates $\tilde{\beta}_{kl}$ are real and independently distributed, the probability that two eigenvalues are equal is almost surely zero on a finite graph.}\textsuperscript{,}\footnote{All contact graphs with homogeneous transition rates do not satisfy Assumption \ref{ass_eig_distinct}. We know that in homogeneous contact graphs, the eigenvalue equals the number of links in the cut set. Consider one infected node and all other nodes susceptible, then the size of the cut set exactly equals the infected node's degree. Since every connected graph has at least two nodes with the same degree \cite{GS}, at least one eigenvalue is degenerate.}
\end{assumption}
%\textsuperscript{,}\footnote{A further implication is that the network must be directed, because for undirected networks, the S-I cut set is equivalent to the I-S cut set.}

Following Assumption \ref{ass_eig_distinct}, the infinitesimal generator $Q$ is diagonalisable and the left- and right eigenvectors from different eigenvalues are orthogonal: $\w_i^T \v_j = \delta_{ij}$ for all $1 \leq i,j \leq 2^N-1$, where $\delta$ indicates the Kronecker delta function \cite{GS}. Then the solution $\s(t)$ in Eq.\ (\ref{eq_MC_general_solution}) simplifies to
\begin{equation}\label{eq_general_solution}
	\s(t) = \pi + \sum_{i=1}^{2^N-1} c_i e^{\lambda_i t} \w_i
\end{equation}
where the constant $c_i = \s(0)^T \v_i$. Moreover, any infinitesimal generator $Q$ has row sum zero and thus a zero eigenvalue with corresponding all-one right-eigenvector $\v_{2^N-1}=\mathbf{u}$. Hence, all left-eigenvectors $\w$ are orthogonal to the right-eigenvector $\v_{2^N-1} = \mathbf{u}$, which implies that each left-eigenvector (except $\w_{2^N-1}$) sums element-wise to zero \cite[art. 140--142]{GS}.

The set of nodes that are initially infected at time $t=0$ remain infected for all times $t>0$, because nodes cannot cure. Consider a configuration $i$ in the Markov graph. If one of the initially infected nodes is susceptible in configuration $i$, then configuration $i$ cannot be reached from the initial state and configuration $i$ can be removed from the Markov graph.

Using Assumption \ref{ass_eig_distinct} and exploiting the upper-triangular structure of the infinitesimal generator $Q$, we can explicitly compute the eigenvectors. We say that state $i$ can \emph{reach} state $j$ if there is a directed path in the Markov graph from state $i$ to state $j$. If state $i$ can reach state $j$, then we denote $i \to j$. 

\begin{figure}[!ht]
	\centering
	\subfloat[\label{fig_example_graph}]{%
		\includegraphics[width=0.45\columnwidth]{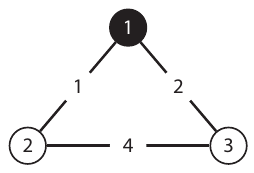}
	}
	\subfloat[\label{fig_example_Markov_graph}]{%
		\includegraphics[width=0.45\columnwidth]{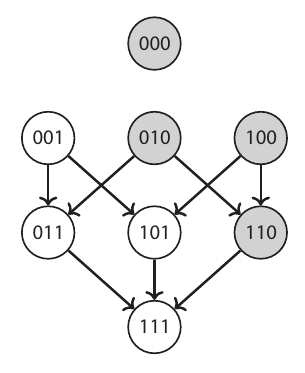}
	}
	\caption{(a) Example of a weighted contact graph with $N=3$ nodes and (b) the corresponding Markov graph. The grey nodes can be removed from the Markov graph, because the SI process starts with node 1 infected (indicated by the black node) and all grey nodes indicate configurations where node 1 is susceptible.}
	\label{fig_example_eigenvectors}
\end{figure}

Let us first consider the right-eigenvectors $\v_i$. For configurations $i$ that correspond to one infected node, its right-eigenvector is the basis vector $\mathbf{e}_i$. For configurations $i$ with two infected nodes, the right-eigenvector will have non-zero elements at the positions that correspond to all states that can reach state $i$ (including state $i$ itself). Let us consider an example with $N=3$ nodes in Figure~\ref{fig_example_eigenvectors}. 

\begin{example}
	The contact graph in Figure~\ref{fig_example_graph} consists of 3 nodes. Node 1 is initially infected whereas node 2 and 3 are susceptible. The Markov graph in Figure \ref{fig_example_Markov_graph} is comprised of only four states: (00), (01), (10) and (11), where $(x_3 x_2)$ represent the viral state of nodes 3 and 2, respectively. We have omitted the viral state of node 1, as node 1 is always infected. The infinitesimal generator~$Q$ equals
	\begin{equation*}
		Q = \begin{pmatrix}
			q_{00} & q_{01} & q_{02} & 0 \\
			0 & q_{11} & 0 & q_{13} \\
			0 & 0 & q_{22} & q_{23} \\
			0 & 0 & 0 & 0
		\end{pmatrix}
	\end{equation*}
	where $q_{00} = -3, q_{01} = 1, q_{02} = 2, q_{11} = -6, q_{13} = 6, q_{22} = -5, q_{23} = 5$. For configuration 0=(00) with eigenvalue $\lambda_0=q_{00}=-3$, the right-eigenvector $\v_0$ equals the basis vector $\mathbf{e}_0 = (1,0,0,0)^T$. For configuration $1=(01)$, the eigenvector can be computed using the definition;
	\begin{align*}
		&(Q - q_{11} I) \v_1  = 0 \quad \Longleftrightarrow \\
		&\quad\left( \begin{array}{cccc|c}   q_{00}-q_{11} & q_{01} & q_{02} & 0 & 0  \\   0 & 0 & 0 & q_{13} & 0  \\   0 & 0 & q_{22}-q_{11} & q_{23} & 0 \\ 0 & 0 & 0 & -q_{11} & 0 \\\end{array} \right)
	\end{align*}
	The last three rows reveal that $(\v_1)_2 = (\v_1)_3 = 0$. Now we choose $(\v_1)_1 = 1$. Then the solution follows as $(\v_1)_0 = \frac{q_{01}}{q_{11} - q_{00}} = -\frac{1}{3}$. We conclude that $\v_1 = (-1/3, 1, 0, 0)^T$ and the only non-zero elements in $\v_1$ correspond to configurations that can reach configuration 1. Analogously, we can find $\v_2 = (-1, 0, 1, 0)^T$. For $\v_3$, we obtain
	\begin{equation*}
		(Q - q_{33} I) \v_3  = 0 \quad \Longleftrightarrow \quad \left( \begin{array}{cccc|c}   q_{00} & q_{01} & q_{02} & 0 & 0  \\   0 & q_{11} & 0 & q_{13} & 0  \\   0 & 0 & q_{22} & q_{23} & 0 \\ 0 & 0 & 0 & 0 & 0 \\\end{array} \right)
	\end{equation*}
	We choose $(\v_3)_3 = 1$. Then the elements $(\v_3)_2 = - \frac{q_{23}}{q_{22}}$ and $(\v_3)_1 = - \frac{q_{13}}{q_{11}}$. The element $(\v_3)_0$ can be computed in the same manner and equals $(\v_3)_0 = -\frac{q_{01} (\v_1)_1 + q_{02} (\v_1)_2}{q_{00}}$. Here, the iterative nature of the construction is clear; $(\v_3)_0$ depends on $(\v_3)_1$ and $(\v_3)_2$. $\hfill\square$
\end{example}

In general, the right-eigenvector $\v_i$ of a certain configuration $i$ can be constructed iteratively;
\begin{align*}
	(\v_i)_j &= 0, \qquad\qquad\text{if state } j \text{ cannot reach state } i \\
	(\v_i)_i &= 1, \qquad\qquad\text{by construction} \\
	(\v_i)_h &= \frac{q_{ih}}{\lambda_i-\lambda_h} = \frac{\displaystyle\sum_{l=1}^N \tilde{\beta}_{ml} x_l(h)}{\lambda_i - \lambda_h}, \\
	&\qquad\qquad\qquad \text{if } i - h = 2^m, m = 0,1,\ldots, N-1 \\
	(\v_i)_g &= \sum_{h=0}^{2^N-1} \frac{q_{hg}}{\lambda_i - \lambda_g} (\v_i)_h \\
	&\qquad\qquad\qquad \text{if } h - g = 2^n, n = 0,1,\ldots, N-1 \\
	\vdots
\end{align*}
Configurations $h$ and $i$ differ only at position $m$, which corresponds to node $m+1$ being infected in configuration~$i$ and susceptible in configuration $h$. Similarly, configuration $g$ and $h$ differ only at position $n$ corresponding to node $n+1$, etc. The iterative formulation consists of at most $N-1$ steps, because at each iteration, one node turns from susceptible into infected, there are $N$ nodes in the contact graph and we start with at least one infected node. We emphasise that this construction is infeasible for degenerate eigenvalues.

The left-eigenvectors $\w_i$ can be derived similarly, from which we find
\begin{align*}
	(\w_i)_h &= 0, \qquad\qquad\text{if state } h \text{ cannot be reached,} \\
	&\qquad\qquad\qquad\qquad\qquad\qquad \text{starting from state } i. \\
	(\w_i)_i &= 1, \qquad\qquad\text{by construction} \\
	(\w_i)_j &= \frac{q_{ij}}{\lambda_i-\lambda_j} = \frac{\displaystyle\sum_{l=1}^N \tilde{\beta}_{ml} x_l(j)}{\lambda_i - \lambda_j}, \\
	&\qquad\qquad\qquad \text{if } j - i = 2^m, m = 0,1,\ldots, N-1 \\
	(\w_i)_k &= \sum_{j=0}^{2^N-1} \frac{q_{jk}}{\lambda_i - \lambda_k} (\w_i)_j \\
	&\qquad\qquad\qquad \text{if } k - j = 2^n, n = 0,1,\ldots, N-1 \\
	\vdots
\end{align*}
By construction, the eigenvectors are orthonormal $\w_i^T \v_i = 1$ for all configurations $i$, because the eigenvectors have only one non-zero value in common, which is the 1 at position $i$. Using the right-eigenvectors $\v_i$ and the left-eigenvectors $\w_i$, the solution (\ref{eq_general_solution}) can be computed explicitly.

The element $(\v_i)_h$ of the right eigenvector $\v_i$ is non-zero if state $h$ can reach state $i$, in other words, if there is a directed path in the Markov graph from state $h$ to state $i$. The opposite holds for the left-eigenvector $\w_i$, whose element $j$ is non-zero if there is a directed path in the Markov graph from state $i$ to state $j$. 

Thus, starting a certain state $i$, the element $h$ of the right-eigenvector $\v_i$ specifies whether the process can start in configuration $h$ and converge to configuration $i$. Similarly, the non-zero element $j$ of the left-eigenvector $\w_i$ specifies whether the process can start in configuration $i$ and converge to configuration $j$.

\subsection{Computational feasibility}
\label{sec_comp_infeasible}
Irrespective of the application domain, the quantity of interest of the SI process should be computable in a reasonable time. The $2^N \times 2^N$ infinitesimal generator $Q$ cannot be computed nor stored if the number of nodes $N \geq 25$. Fortunately, the exact solution $\s(t)$ under Assumption~\ref{ass_eig_distinct} can be computed without explicitly constructing the infinitesimal generator $Q$. 

Only the eigenvalues and eigenvectors of the infinitesimal generator $Q$ are required in the solution (\ref{eq_general_solution}). For a given configuration $i$, the eigenvalue $\lambda_i$ can be computed based on equation (\ref{eq_eigenvalues2}) in $\O(N^2)$ operations. To compute all $2^N$ eigenvalues, we need $\O(N^2 2^N)$ operations. The iterative procedure to compute a single left- and right-eigenvector takes $\O(N 2^N)$ operations if the eigenvalues are known. Since there are $2^N$ eigenvectors, the total number of required operations to compute all eigenvectors is $\O(N 4^N)$. Hence, for networks with $N>20$ nodes, computing all eigenvalues and eigenvectors remains infeasible.

Even though the computation of all eigenvalues and eigenvectors is infeasible for large networks, for certain quantities of interest, not all eigenvalues or eigenvectors are required. Here, we consider two exemplary scenarios. First, we consider the probability $\P[X_2(t) = 1]$ that node 2 is infected at time $t$. To compute $\P[X_2(t)=1]$, we sum over all states in $\mathbf{S}$ in which node 2 is infected. We introduce the $2^N \times 1$ vector $\mathbf{m} = (0,1,1,0,0, 1,\ldots,1)^T$, which is $1$ at position $j$ if the binary representation of configuration $j$ has a 1 at position 2, and zero otherwise. Then $\P[X_2(t)=1] = \mathbf{m}^T s(t)$ and using (\ref{eq_general_solution}), we find
\begin{equation}\label{eq_solution_X2}
	\P[X_2(t)=1] = \mathbf{m}^T \pi + \sum_{i=1}^{2^N-2} c_i e^{\lambda_i t} \mathbf{m}^T \w_i.
\end{equation}
The inner product $\mathbf{m}^T \pi = 1$, because node 2 will be infected after infinitely long time. We know that each left-eigenvector $\w_i$ sums element-wise to zero. Then for all configurations $i$ for which the non-zero elements of the left-eigenvector $\w_i$ overlap with the non-zero elements of $\mathbf{m}$, the inner product is zero. Only the states $i$ that can be reached from state $2$ will have a zero inner product, which are $2^{N-1}$ states. Since $c_i$ is non-zero for $2^{N-1}$ states, we conclude that the sum in (\ref{eq_solution_X2}) can be simplified from $2^N-1$ states to $2^{N-2}$ states. Unfortunately, summing over $2^{N-2}$ states is still exponentially large and solving for large networks remains impossible.

As a second example, we consider the probability that all nodes are infected at time $t$. We multiply the solution $\s(t)$ with the vector $\mathbf{m} = (0, 0, \ldots, 0, 1)^T$, which is the all-zeros vector, except the last element is one. Regarding the product $m^T \w_i$ in \eqref{eq_solution_X2}, we conclude that only the last element $(\w_i)_{2^N-1}$ is of importance. Unfortunately, computing the last element of the left-eigenvector $\w_{i}$ is as difficult as computing the whole vector, because of the iterative construction of the eigenvector $\w_{i}$.

Another key property of epidemics involves the average fraction of infected nodes, also known as the {\em prevalence}. Computation of the prevalence requires the computation of the probability of infection of each node individually, which is definitely computationally harder than the previous two examples.

The provided examples consider limit cases, namely the infection probability of a single node and all nodes, respectively. For both examples, exponentially many eigenvalues and eigenvectors are required to build up the solution (\ref{eq_general_solution}). The examples illustrate that calculating the exact, time-varying solution of the infection probability in SI epidemics on large contact graphs is generally infeasible.

\subsection{Numerical simulations}\label{sec_simulations}
\begin{figure*}[!ht]
	\centering
	\subfloat[\label{fig_SI_ER}]{%
		\includegraphics[width=0.48\textwidth]{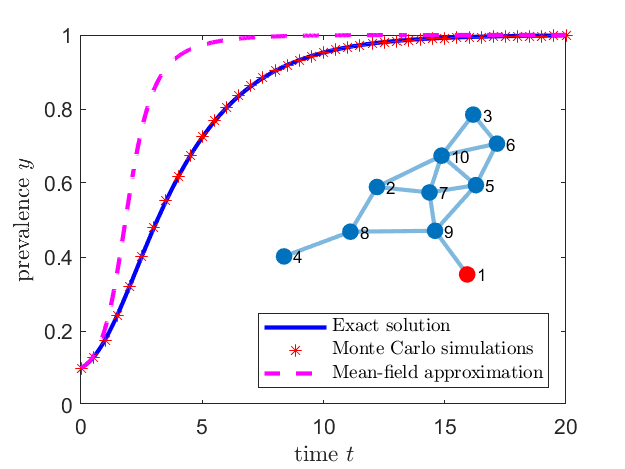}
	}
	\subfloat[\label{fig_SI_cycle}]{%
		\includegraphics[width=0.48\textwidth]{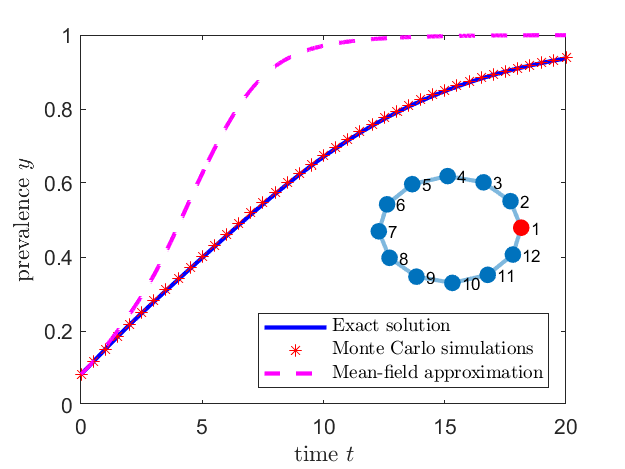}
	}
	\caption{The exact solution (solid line), Monte Carlo simulations (asterisks) and the mean-field approximation (dashed line) of the Markovian SI process on (a) an \ER{} graph with $N=10$ nodes and $p=0.33$ and (b) a cycle graph with $N=12$ nodes. The infection rates $\beta_{ij}$ are chosen uniformly at random between $\beta_{\min} = 0.1$ and $\beta_{\max}=1.1$. Initially, node 1 is infected. The mean-field approximation deviates in both cases significantly from the exact solution.}
	\label{fig_SI_examples}
\end{figure*}

We illustrate the accuracy of our exact solution method in Figure \ref{fig_SI_examples} for two contact graphs: an \ER{} graph with $N=10$ nodes and link-connectivity $p=0.33$ and a cycle graph with $N=12$ nodes. The simulations start with node 1 infected. The results are averaged over 10,000 Monte Carlo simulations. We compare our exact solution and the Monte Carlo simulations with the N-Intertwined Mean-Field Approximation (NIMFA), which assumes every pair of random variables is uncorrelated \cite{IntroNIMFA}. The exact solution coincides with the Monte Carlo simulations, whereas the mean-field approximation vastly overestimates the time-dependent prevalence \cite{PositivelyCorrelatedSIS}.

\begin{figure}[!ht]
	\centering
	\includegraphics[width=0.98\columnwidth]{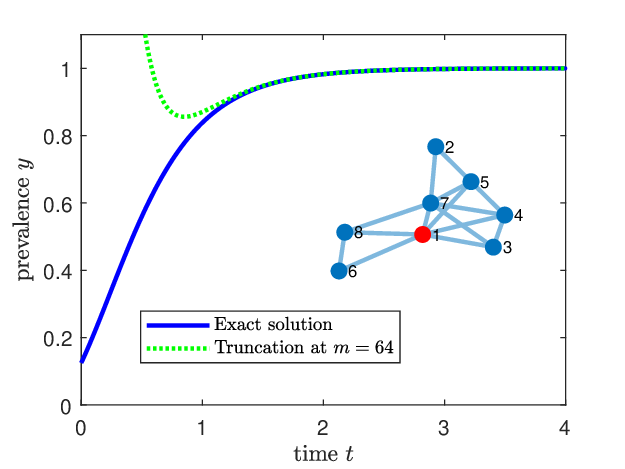}
	\caption{The exact solution (solid line) and the truncated solution (dotted line) with $m=N^2$ out of $2^N$ states of the Markovian SI process on an \ER{} graph with $N=8$ nodes. The infection rates~$\beta_{ij}$ are chosen uniformly at random between $\beta_{\min} = 0.9$ and $\beta_{\max}=1.1$. Initially, node 1 is infected.}
	\label{fig_truncation_method}
\end{figure}

As we concluded earlier, the large (in modulus) eigenvalues have a small effect on the dynamics at large times. Presumably, the exact solution (\ref{eq_general_solution}) can be accurately approximated using a truncation method. The number of modes is truncated at index~$m$, such that the approximated solution $\tilde{\s}$ equals
\begin{equation}\label{eq_solution_truncated}
	\tilde{\s}(t) \approx \pi + \sum_{i=1}^{m} c_i e^{\lambda_i t} \w_i
\end{equation}
The approximation of Eq.\ (\ref{eq_general_solution}) by Eq.\ (\ref{eq_solution_truncated}) introduces the error
\begin{equation*}
	e(t) = \| \s(t) - \tilde{s}(t) \| = \left\| \sum_{i=m+1}^{2^{N-1}} c_i e^{\lambda_i t} \w_i \right\| \leq \sum_{i=k+1}^{2^{N-1}} e^{\lambda_i t} \| c_i \w_i \|
\end{equation*}
where $\| \cdot \|$ is a vector norm and we used the triangle inequality $\| \mathbf{a} + \mathbf{b} \| \leq \| \mathbf{a} \| + \| \mathbf{b} \|$. Using the fact that the eigenvectors are normalised, we find
\begin{equation*}
	e(t) \leq \sum_{i=m+1}^{2^{N-1}} e^{\lambda_i t} \| c_i \w_i \| < e^{\lambda_{m+1} t} \sum_{i=m+1}^{2^{N-1}} | c_i |
\end{equation*}
such that the error scales as $\O(e^{\lambda_{m+1} t})$. Similarly to $\eps$-SIS dynamics \cite[Appendix D]{AchterbergMieghemSISeigenvalues}, the truncation method performs poor for small times, see Figure~\ref{fig_truncation_method}. The main reason is as follows. Suppose the SI epidemic starts at a single node that has only one link with a very small weight. Under the truncation approximation, the eigenmode corresponding to the infection from the seed node to the second node will be disregarded, whereas this link has a profound impact on the dynamics.

\subsection{Extensions of Markovian SI epidemics}\label{sec_extensions}
For Markovian SI processes on heterogeneous networks, we have derived the explicit solution (\ref{eq_general_solution}). To demonstrate the versatility of our method, we extend our results on the SI process to non-Markovian epidemics, temporal networks, simplicial contagion and the inclusion of self-infections. %Additionally, we briefly discuss the difficulties of computing the exact solution for general compartmental models.

\subsubsection{Non-Markovian dynamics}
\begin{figure}[!ht]
	\centering
	\includegraphics[width=0.98\columnwidth]{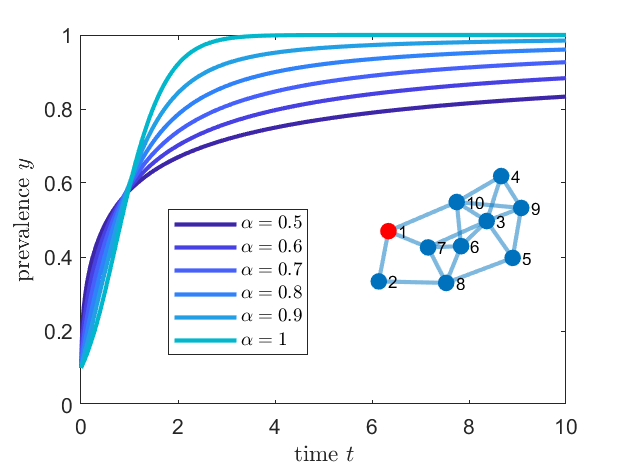}
	\caption{The exact solution of the non-Markovian SI process with Caputo fractional derivative $\alpha$ on an \ER{} graph with $N=10$ nodes. The infection rates~$\beta_{ij}$ are chosen uniformly at random between $\beta_{\min} = 0.9$ and $\beta_{\max}=1.1$. Initially, node 1 is infected. The case $\alpha=1$ coincides with the Markovian case.}
	\label{fig_nonmarkov}
\end{figure}

Most epidemic models assume a memoryless Markov process, i.e.\ the probability to infect a neighbour is exponentially distributed over time. Non-Markovian effects in Markov chains can be taken into account using fractional calculus, as was recently derived by Van Mieghem \cite{vanmieghem2022fractionalSIS}. The governing equations (\ref{eq_MC_diff}) of the Markov chain change to
\begin{equation}\label{eq_MC_alpha}
	D^\alpha \s_\alpha^T(t) = \s_\alpha^T(t) Q^\alpha,
\end{equation}
where $D^\alpha$ is the Caputo fractional derivative and $0 < \alpha < 1$. As derived by Van Mieghem \cite{vanmieghem2022fractionalSIS}, the solution (\ref{eq_general_solution}) becomes
\begin{equation}\label{eq_general_solution_alpha}
	\s(t) = \pi + \sum_{i=1}^{2^N-1} c_i E_{\alpha,1} \left( (\lambda_i t)^\alpha \right) \w_i
\end{equation}
where $E_{\alpha,1}(z)$ is the Mittag-Leffler function \cite{vanmieghem2021mittagleffler} of the complex number $z$. Up to our knowledge, Eq.\ (\ref{eq_general_solution_alpha}) is the first exact solution of a non-Markovian epidemic process on an heterogeneous network. The exact solution (\ref{eq_general_solution_alpha}) is plotted in Figure~\ref{fig_nonmarkov} for $0.5 \leq \alpha \leq 1$. Decreasing~$\alpha$ increases the prevalence at short times but slows down the disease spread for large times. Compared to the Markovian case $\alpha=1$, decreasing $\alpha$ increases the variance of the infection time distribution, causing a large probability of both very slow and very fast infections. At short times, this results in a larger infection probability, but it also takes much longer for all nodes to become infected.

\subsubsection{Temporal networks}
So far, we have assumed that the contact network is fixed. In reality, contact networks are varying over time due to movements of individuals. Given a temporal contact graph $G$ that changes finitely many times in the interval $[0,T]$, each connected contact graph $G_n$ in an interval $[t_n, t_{n+1}]$ is represented by the adjacency matrix $A^{(n)}$. We assume that the contact graph remains connected at all times. Then we may compute the eigenvectors and eigenvalues for each of those adjacency matrices $A^{(n)}$ and the solution becomes
\begin{equation}\label{eq_general_solution_temporal}
	\s(t) = \pi + \sum_{i=1}^{2^N-1} c_i^{(n)} e^{\lambda_i^{(n)} t} \w_i^{(n)}, \qquad t_n \leq t \leq t_{n+1}
\end{equation}
where $c_i^{(n)} = \s(t_n) \v_i^{(n)}$. Within each interval $t_n \leq t \leq t_{n+1}$, the dynamics are equivalent to the case of the fixed network. Going to the next time interval requires different eigenvalues and eigenvectors, and additionally, the initial condition of the new interval must equal the final state of the previous interval. Although the eigenvectors and eigenvalues may be different for each time interval, each contact graph is assumed to be connected, thus the steady-state vector~$\pi=(0,0,\ldots,0,1)^T$, representing the state in which all nodes are infected.

\subsubsection{Simplicial contagion}
In the study of opinion dynamics on networks, it is observed that multiple neighbours of a node may be required to persuade a node to adopt a particular strategy or opinion. Such higher-order spreading phenomena on networks are known as simplicial contagion \cite{centola2007complexcontagion}, which was first applied to the SIS model by Iacopini \emph{et al.} \cite{DriehoekSIS}. The conceptual idea is that, besides node-node interactions, connected infected triangles may increase the probability of converting a particular node to the other opinion, even larger than the sum of the individual infection rates. Adding simplicial contagion to the SI process does not alter the \emph{structure} of the infinitesimal generator $Q$, because simplicial infections require the existence of triangles in the original network structure, whose individual links already had contributions to the infinitesimal generator. Instead, the \emph{link weights} corresponding to triangles are increased for certain configurations in the infinitesimal generator $Q$. The eigenvalues are no longer the weighted sum over all links in the S-I cut set, but additionally include the weighted sum over all 2-infected-1-susceptible-node triangles in the contact graph. The eigenvectors can be computed in the usual way, such that the time-dependent solution (\ref{eq_general_solution}) can be obtained.

\begin{figure}[!ht]
	\centering
	\includegraphics[width=0.98\columnwidth]{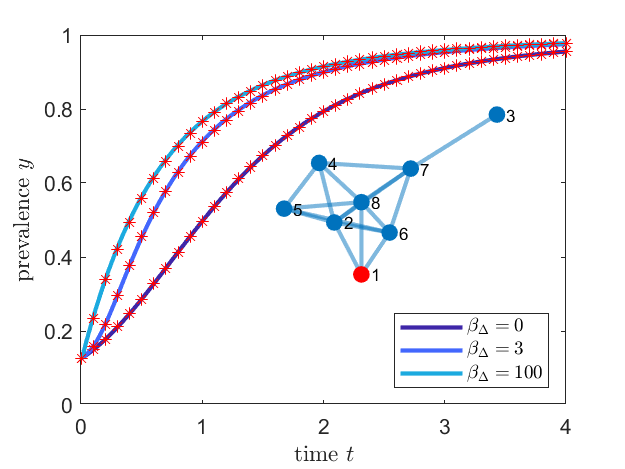}
	\caption{The exact solution of the SI process with simplicial contagion on an \ER{} graph with $N=8$ nodes. The infection rates~$\beta_{ij}$ are chosen uniformly at random between $\beta_{\min} = 0.2$ and $\beta_{\max}=1.0$ and the triangle interaction rate $\beta_\Delta$ is taken equal to $0, 3$ and $100$. Initially, node 1 is infected. The Monte Carlo simulations are shown as red asterisks.}
	\label{fig_example_simplicial_contagion}
\end{figure}

Figure~\ref{fig_example_simplicial_contagion} illustrates the impact of simplicial contagion on the classical SI process. Adding the homogeneous simplicial infection rate $\beta_\Delta$ increases the prevalence $y$, but only up to a certain point, because some parts of the contact graph do not contain any triangles and therefore nodes cannot be infected by higher-order simplexes.

\subsubsection{Self-infections}
Contrary to the standard SI process, where the spread or discovery of items is solely related to the contact graph, some spreading phenomena can be triggered by external processes, which are unrelated to the spread on the contact graph. For example, individuals or companies may adopt a certain product or technology without any interference with others. The situation in which a node becomes infected spontaneously without inference with other nodes, is described as a self-infection with rate~$\eps$. The Bass model describes the spread of direct infections and self-infections of new products used in companies \cite{bassmodel} and is actually equivalent to the SI process with self-infections \cite{IntroductionEpsSIS}.

The infinitesimal generator $Q$ of the $\eps$-SI process equals
\begin{align}
	q_{ij} &= \eps_m + \sum_{k=1}^N \tilde{\beta}_{mk} x_k(i), \qquad \text{if } j = i + 2^{m-1} \\
	&\qquad\qquad \text{ with } m = 1,2,\ldots, N \text{ and } x_m(i)=0. \nonumber \\
	q_{ii} &= - \sum_{\substack{k=0 \\ k\neq i}}^{2^N-1} q_{ki},
\end{align}
Thus, adding self-infections can both increase the link weight and add more links in the Markov graph. The resulting infinitesimal generator $Q$ will remain upper-triangular, because the total number of infected nodes can only increase. In this case, the eigenvalue $\lambda_i$ equals the weighted sum of the links in the S-I cut set plus the total (possibly weighted) sum of the self-infection rate of all susceptible nodes in configuration~$i$.

\begin{figure}[!ht]
	\centering
	\includegraphics[width=0.98\columnwidth]{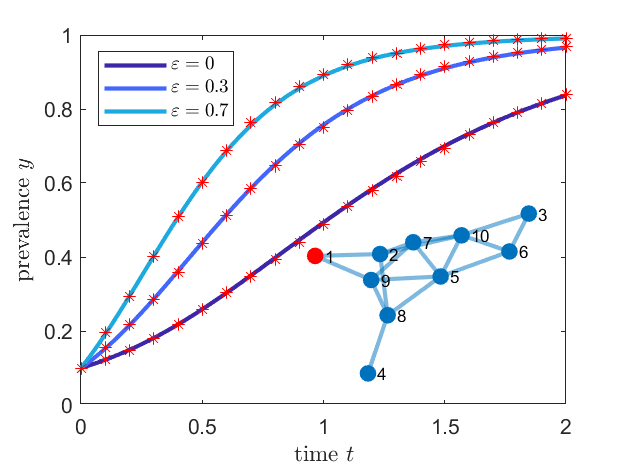}
	\caption{The exact solution of the Markovian SI process with self-infections on an \ER{} graph with $N=10$ nodes. The infection rates~$\beta_{ij}$ are chosen uniformly at random between $\beta_{\min} = 0.9$ and $\beta_{\max}=1.1$. The self-infection rate $\eps$ is taken equal to $0, 0.3$ and $0.7$. Initially, node 1 is infected.}
	\label{fig_example_epsSI}
\end{figure}

Adding the self-infection process to the SI process increases the prevalence significantly, as shown in Figure~\ref{fig_example_epsSI}. In the limit where the self-infection process is much larger than the infection process, each node adopts the innovation independent of all other nodes and the prevalence $y$ is just the sum of $N$ independent exponential random variables with rate $\eps$.

\begin{figure*}[!htb]
    \centering
    \includegraphics[width=0.67\textwidth]{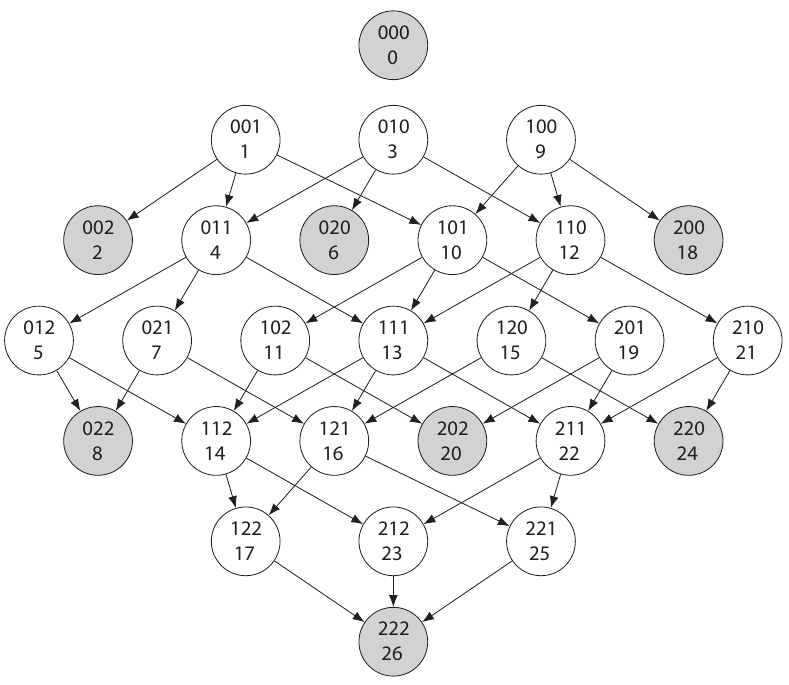}
    \caption{The Markov graph of the SIR process on the complete graph with $N=3$ nodes. Each state $(x_3 x_2 x_1)$ encodes the configuration (viral state) of the nodes; susceptible (0), infected (1) and recovered (2). The configuration number $i$ is shown in bold.}
    \label{fig_markov_graph_SIR}
\end{figure*}

\section{The SIR process}\label{sec_SIR}
Contrary to the SI process, whose applicability is limited, network-based SIR epidemics has a very rich and long history in modelling network epidemiology \cite{ReviewPaperSIS}. The SIR process is a transient process, because after being infected, nodes will recover and can never be re-infected. We solve here the SIR process on a heterogeneous network by proposing a labelling of the state space based on trinary numerals.

Again, to simplify the notation, we denote the $N \times N$, possibly asymmetric, weighted and non-negative adjacency matrix $B$ with elements $\tilde{\beta}_{kl} = a_{kl} \beta_{kl}$.

We introduce $x_k$ as the viral state of node $k$: 
\begin{align*}
	x_k &= 0, \qquad \text{if node } k \text{ is susceptible} \\
	x_k &= 1, \qquad \text{if node } k \text{ is infected} \\
	x_k &= 2, \qquad \text{if node } k \text{ is recovered}
\end{align*}
The viral state vector $\mathbf{x} = (x_N, x_{N-1}, \ldots, x_1)^T$ describes the viral state of all nodes. Given a particular viral state vector~$\mathbf{x}$, we may compute the configuration number $i$ using the trinary numerals;
\begin{equation*}
	i = \sum_{k=1}^N x_k(i) \cdot 3^{k-1}
\end{equation*}
The trinary numbering ensures that each viral state vector $\mathbf{x}$ corresponds to a unique configuration number $i$, which ranges between $i=0$ and $i=3^{N}-1$. Then the infinitesimal generator $Q$ of the SIR process equals
\begin{align*}
	q_{ij} &= \sum_{k=1}^N \tilde{\beta}_{mk} \cdot \mathbf{1}_{\{ x_k(i)=1 \}}, \qquad \text{if } j=i+3^{m-1} \\
	&\qquad\qquad\text{ with } m=1,2,\ldots,N \text{ and } x_m(i)=0 \nonumber \\
	q_{ij} &= \delta_m, \qquad\qquad \text{if } j=i+3^{m-1} \\
	&\qquad\qquad\text{ with } m=1,2,\ldots,N \text{ and } x_m(i)=1 \nonumber \\
	q_{ii} &= - \sum_{\substack{j=0 \\ j\neq i}}^{3^N-1} q_{ji},
\end{align*}
where $\mathbf{1}_{\{x\}}$ is the indicator function, which is one if condition $x$ is satisfied and zero otherwise. The Markov graph specifies all possible transitions between the $3^N$ states and the rates are given by the infinitesimal generator $Q$. The Markov graph for a complete graph with $N=3$ nodes is shown in Figure~\ref{fig_markov_graph_SIR}. The absorbing states are indicated by shaded circles. We observe that the Markov graph is a directed tree, i.e.\ a bipartite graph. Moreover, we can define the \emph{layer number} $l(i) = I(i) + 2R(i)$, where $I(i)$ represents the number of infected nodes and $R(i)$ the number of recovered nodes in configuration $i$. Transitions can only take place from states in layer $l$ to states in layer $l+1$. Layers with odd number $l$ do not contain any absorbing states, because an odd layer number $l$ implies that at least one node must be infected. If a configuration contains one infected node, that node can always recover, ruling out the existence of absorbing states in that layer.

\begin{figure}[!ht]
	\centering
	\includegraphics[width=0.9\columnwidth]{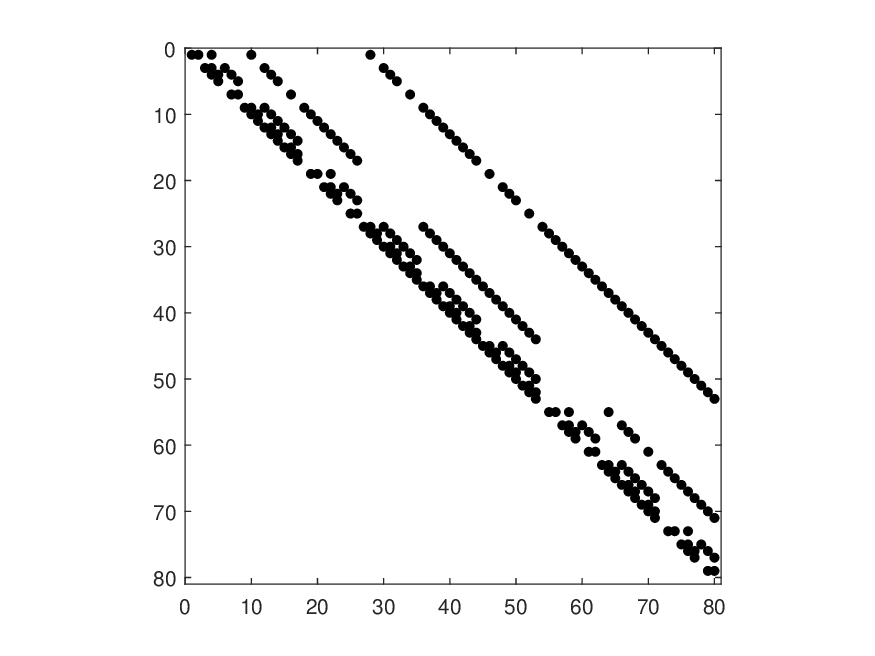}
	\caption{The structure of the infinitesimal generator $Q$ for the SIR process on a complete graph with $N=4$ nodes.}
	\label{fig_Q_SIR}
\end{figure}

Figure~\ref{fig_Q_SIR} shows all non-zero elements of the infinitesimal generator~$Q$ with $N=4$ nodes. Some rows of $Q$ are empty, because these rows correspond to absorbing states in the SIR process (the grey nodes in Figure~\ref{fig_markov_graph_SIR}). Although the infinitesimal generator $Q$ has dimensions $3^N \times 3^N$, the total number of transitions is much lower. For a given configuration $i$, the number of possible transitions is given by $N$, because each node can change its viral state to at most one other state. Taking into account the possibility of not making any change, the number of non-zero elements in the infinitesimal generator $Q$ is upper bounded by $(N + 1) 3^N$, which is much less than a dense matrix structure with $3^{2N}$ elements.

\subsection{SIR eigenvalues}\label{sec_SIR_eigenvalues}
Similar to the SI process, the eigenvalue~$\lambda_i$ of the SIR process can be understood to belong to configuration~$i$ because the infinitesimal generator~$Q$ is triangular. The eigenvalue~$\lambda_i$ of configuration~$i$ equals:
\begin{equation}\label{eq_SIR_eig}
	\lambda_i = - \sum_{k=1}^N \sum_{l=1}^N \tilde{\beta}_{kl} \cdot \mathbf{1}_{\{ x_k(i)=1 \cap \, x_l(i)=0 \}} - \sum_{k=1}^N \delta_k \cdot \mathbf{1}_{\{ x_k(i)=1 \}}
\end{equation}
The first component specifies the sum over the weighted links in the S-I cut set as in \eqref{eq_eigenvalues2}, whereas the second part contains the total weighted curing rate of the infected nodes in configuration $i$. Each configuration $i$ that does not contain any infected nodes (i.e. $\mathbf{1}_{\{ x_k(i)=1 \}} = 0$ for all $k$) is an absorbing state and cannot be left. The total number of states without infected nodes, thus only containing susceptible and recovered nodes, equals $2^N$. Hence, there are $2^N$ absorbing states. The eigenvalue of an absorbing state equals $\lambda=0$ and eigenvalue $\lambda=0$ has algebraic multiplicity $2^N$. To determine the geometric multiplicity of eigenvalue $\lambda=0$, i.e.\ solving $Q \mathbf{v} = \mathbf{0}$, we investigate the structure of the infinitesimal generator $Q$ in Figure~\ref{fig_Q_SIR}. There are $2^N$ rows with only zeros, allowing for $2^N$ free variables in the right-eigenvector $\mathbf{v}$. Hence, the algebraic multiplicity is equal to the geometric multiplicity for $\lambda=0$ and moreover, the eigenvectors with eigenvalue zero can be chosen orthogonally.

\begin{figure}[!ht]
	\centering
	\includegraphics[width=0.92\columnwidth]{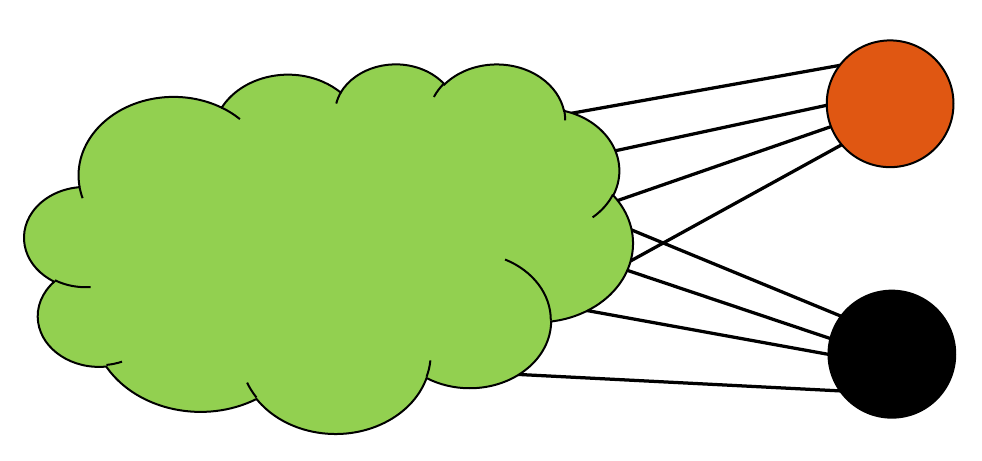}
	\caption{The contact graph $G$ is represented by a green cloud containing susceptible nodes, an infected black node and the orange node. The link between the black and orange node is missing. Whether the orange node is susceptible or recovered does not influence the computation of the eigenvalue $\lambda$ from Eq.\ \eqref{eq_SIR_eig}, thus the eigenvalue $\lambda$ is degenerate.}
	\label{fig_SIR_noncomplete}
\end{figure}

It remains to verify the uniqueness of the non-zero eigenvalues. Similar as for SI epidemics, we require that all infection rates $\tilde{\beta}_{kl}$ are different. For SIR epidemics, we additionally require that the contact graph is complete. Consider the non-complete contact graph $G$ from Figure~\ref{fig_SIR_noncomplete} with a missing link between the orange and black node. Suppose the black node is infected and the orange node is either susceptible or recovered. All other nodes are considered susceptible (represented by the cloud). The eigenvalue $\lambda$ in \eqref{eq_SIR_eig} equals the SI cut set plus the curing rate of the black node. The fact that the orange node is susceptible or recovered does not influence the eigenvalue $\lambda$, because the SI cut set and the sum over the curing rates are equivalent in both cases. Thus, any non-complete contact graph has one or more degenerate non-zero eigenvalues.

From simulations it appears that even in the case of missing links in the contact graph (and subsequently degenerate eigenvalues), the corresponding eigenspace is of full rank and orthogonal eigenvectors can be found. Explicit constructions for those eigenvectors are cumbersome and are omitted here. Instead, we can set infection rates $\tilde{\beta}_{kl}>0$ arbitrarily small for every non-existing link. The resulting graph is the complete graph with heterogeneous infection rates, which can accurately approximate any realistic situation at the cost of a small error and avoids the computation of the complicated case of degenerate eigenvalues.

\subsection{SIR eigenvectors}\label{sec_SIR_eigenvectors}
The right- and left-eigenvectors $\v_i$ and $\w_i$ of the SIR process can be constructed in a similar fashion as for the SI process in Section~\ref{sec_exact_solution}. We focus on the right-eigenvector $\v_i$, but the left-eigenvector $\w_i$ can be constructed similarly. 

For a configuration $i$ that corresponds to one infected node, the right-eigenvector $\v_i$ is the basis vector $\mathbf{e}_i$. For a configuration~$i$ with two infected nodes, the right-eigenvector $\v_i$ will have non-zero elements at the positions that correspond to all states that can reach state $i$ (including state $i$ itself). In fact, we follow the same procedure as for the SI process. The only exception occurs when $\lambda_i=0$, which is a degenerate eigenvalue. Fortunately, for $\lambda=0$ the algebraic and geometric multiplicity are equal, such that the eigenspace is of full rank and we may choose the eigenvectors of the zero eigenvalues as standard basis vectors: for configuration $i$ with eigenvalue $\lambda_i=0$, we choose $\v_i = \mathbf{e}_i$. 

In general, the right-eigenvector $\v_i$ of a certain configuration $i$ can be constructed iteratively;
\begin{align*}
	\textbf{If } &\mathbf{\lambda_i = 0;} \textbf{ then } \v_i = \mathbf{e}_i \\
	\textbf{If } &\mathbf{\lambda_i \neq 0;} \textbf{ then} \\
	&(\v_i)_j = 0, \qquad\qquad\text{if state } j \text{ cannot reach state } i \\
	&(\v_i)_i = 1, \qquad\qquad\text{by construction} \\
	&(\v_i)_h = \frac{q_{ih}}{\lambda_i-\lambda_h} \\
	&\phantom{(\v_i)_h}= \frac{\displaystyle\sum_{l=1}^N \tilde{\beta}_{ml} \cdot \mathbf{1}_{\{ x_l(h)=1 \cap \, x_m(h)=0\}} + \delta_m \cdot \mathbf{1}_{\{ x_m(h)=1\}} }{\lambda_i - \lambda_h} \\ % \cdot \mathbf{1}_{\{\mathbf{x}_m(h)=1\}}
	&\qquad\qquad\qquad \text{if } i - h = 3^m, m = 0,1,\ldots, N-1 \\
	&(\v_i)_g = \sum_{h=0}^{3^N-1} \frac{q_{hg}}{\lambda_i - \lambda_g} (\v_i)_h \\
	&\qquad\qquad\qquad \text{if } h - g = 3^n, n = 0,1,\ldots, N-1 \\
	&\hdots
\end{align*}
Configurations $h$ and $i$ differ only at position $m$ (corresponding to node $m+1$), where node $m+1$ is infected in configuration $i$ and susceptible in configuration $h$ or node $m+1$ is recovered in $i$ and infected in $h$. Similarly, configuration $g$ and $h$ differ only at position $n$, corresponding to node $n$, etc.

\subsection{SIR solution}\label{sec_SIR_solution}
After the derivation of the orthonormal eigenvectors for the zero and non-zero eigenvalues, the infinitesimal generator $Q$ is diagonalisable and the time-dependent solution $\s(t)$ becomes
\begin{equation}\label{eq_solution_SIR}
	\s(t) = \sum_{i=0}^{3^N-1} c_i e^{\lambda_i t} \w_i
\end{equation}
where $c_i = \s(0)^T \v_i$ and $\v_i, \w_i$ are the right and left-eigenvector of $Q$, respectively. Defining the $3^N \times 1$ vector $\mathbf{m}$, whose elements $m_i$ equal the number of infected nodes in configuration $i$, the time-varying prevalence $y(t)$ follows as
\begin{equation*}
	y(t) = \frac{1}{N} \sum_{i=0}^{3^N-1} c_i e^{\lambda_i t} \mathbf{m}^T \w_i
\end{equation*}
which can be simplified to
\begin{equation}\label{eq_solution_SIR_y}
	y(t) = \sum_{i=0}^{2^N-1} \tilde{c}_i + \sum_{i=0}^{3^N-2^N-1} \tilde{c}_i e^{\lambda_i t}
\end{equation}
where $\tilde{c}_i = \frac{1}{N} c_i \mathbf{m}^T \w_i$. The first term in Eq.\ \eqref{eq_solution_SIR_y} contains all zero eigenvalues and the second term contains all non-zero (negative) eigenvalues. Surprisingly, the time-dependent prevalence $y(t)$ in \eqref{eq_solution_SIR_y} is just a function of the eigenvalues $\lambda_i$ and the (complicated) variables $\tilde{c}_i$. 

\begin{figure}[!ht]
	\centering
	\includegraphics[width=0.98\columnwidth]{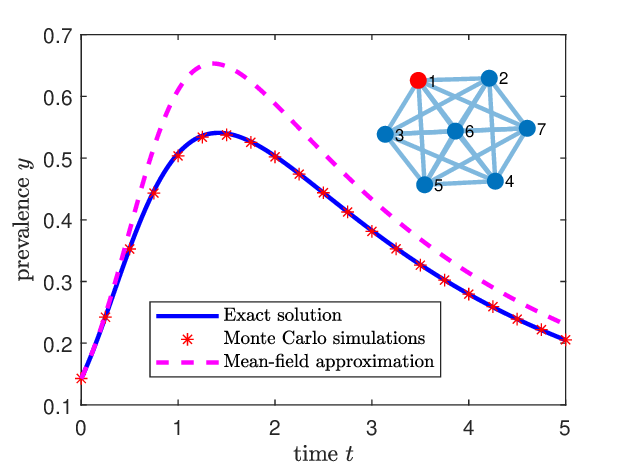}
	\caption{The exact solution of the Markovian SIR process on a complete graph with $N=7$ nodes. The infection rates~$\tilde{\beta}_{kl}$ are chosen uniformly at random between $\beta_{\min} = 0.2$ and $\beta_{\max}=1.0$ and the curing rates $\delta_k$ are chosen uniformly between $\delta_{\min}=0.1$ and $\delta_{\max} = 0.6$. Initially, node 1 is infected.}
	\label{fig_example_SIR}
\end{figure}

We can further simplify \eqref{eq_solution_SIR_y}. The contribution of all non-zero eigenvalues in the second term of \eqref{eq_solution_SIR_y} converges exponentially fast to zero for $t\to\infty$, whereas the first term in \eqref{eq_solution_SIR_y} corresponding to the zero eigenvalues, is fixed. Since the prevalence $y\to0$ for $t\to\infty$, it follows that the first term must be zero. Hence, the solution \eqref{eq_solution_SIR_y} reduces to
\begin{equation}\label{eq_solution_SIR_y2}
	y(t) = \sum_{i=0}^{3^N-2^N-1} \tilde{c}_i e^{\lambda_i t}
\end{equation}
Figure~\ref{fig_example_SIR} shows the exact solution (\ref{eq_solution_SIR_y2}) and Monte Carlo simulations, which match perfectly. For SI epidemics, we considered non-Markovian dynamics using fractional derivatives, included time-varying contact networks, added higher-order simplicial contagion and added self-infections. The same extensions can straightforwardly be applied to SIR epidemics.

\subsubsection{Explicit SIR solution}
Since the Markov graph only contains directed links and re-infections are impossible, the SIR process is transient. Written as a system of $3^N$ linear differential equations, the equations \eqref{eq_MC_diff} can be solved exactly \cite{vanmieghem2023lineardiff}. Assuming that the initial condition $\s(0)=0$, except for some state~$i$ where $s_i(0)=1$, then the exact solution follows as
\begin{align}
\begin{split}
    s_j(t) &= 0, \qquad \text{if } j < i \\
    s_i(t) &= e^{\lambda_i t}, \\
    s_j(t) &= q_{jl}\left(
    \frac{e^{\lambda_j t}-e^{\lambda_l t}}{\lambda_l -\lambda_j} \right) - \sum_{h=2}^{l-j} (-1)^{h} \sum_{k_{1}=j+1}^{l-1} \sum_{k_{2}=k_{1}+1}^{l-1} \cdots \\
    &\quad \sum_{k_{h-1}=k_{h-2}+1}^{l-1} q_{j,k_{1}} \left(    {\prod_{r=1}^{h-2}} q_{k_{r},k_{r+1}}\right)  q_{k_{h-1},l}\\
    & \hspace{0.5cm} \times\sum_{n=0}^{h}\frac{e^{\lambda_{k_n} t}}{ {\prod\nolimits_{r=0;r\neq n}^{h}} \left(  \lambda_{k_r} -\lambda_{k_n} \right) }, \qquad \text{if } j > i
    \end{split}
    \label{eq_SIR_exact_whole_formula}
\end{align}
where $h$ denotes the number of hops of a path between starting state $i$ and final state $j$. In fact, Eq.\ \eqref{eq_SIR_exact_whole_formula} is just the non-vectorised form of \eqref{eq_solution_SIR}, directly providing intuition for the coefficients $\tilde{c}_i$ and left-eigenvectors $\w_i$. 

The maximum hopcount $h=2N-1$ in Eq.\ \eqref{eq_SIR_exact_whole_formula}, because of the following reason. Consider the most extreme case, where the process starts with a single infected node and all other nodes are susceptible and the process ends with all nodes recovered. During the process, $N-1$ infections and $N$ recoveries must take place, leading to a total of $2N-1$ transitions. The path is denoted by $\mathcal{P} = (i, k_1, k_2, \ldots, k_{h-1}, j)$. Naturally, not all paths are valid paths, because of the physical constraints of the SIR process (i.e.\ nodes must be infected before they can recover). In Appendix~\ref{app_SIR_Markov_graph}, we demonstrate that the maximal number of paths $\mathcal{P}$ is at least $N!$ for any connected contact graph $G$. Thus, in spite of having the closed-form analytical solution \eqref{eq_SIR_exact_whole_formula}, it is infeasible to compute the exact solution for any moderately sized network.

Surprisingly, the Markovian SIR process with homogeneous and heterogeneous infection and curing rates can be solved exactly on tree graphs with initially one infected node \cite{sharkey2015exactSIRtrees,hall2023exactSIRtrees}. The paradoxical discrepancy is due to our computation of the exact solution for all states, whereas \cite{sharkey2015exactSIRtrees,hall2023exactSIRtrees} merely focus on global properties of the SIR dynamics, such as the prevalence and the average fraction of recovered nodes.

\subsection{Epidemic peak time in SIR}\label{sec_SIR_peak}
A key property of SIR epidemics is the \emph{epidemic peak time}, which is the time at which on average most nodes are infected. Knowing when the number of infections reaches the peak allows decision-makers to employ timely counter-measures. If the infection rate in the SIR process is above the epidemic threshold, then an outbreak occurs with large probability. After potentially a long time, the disease dies out, exhibiting a peak in the number of cases at the epidemic peak time. Due to the heterogeneous infection and curing rates, multiple local extrema may be observed \cite{alutto2022SIRpeaks}.

The epidemic peak can be derived explicitly for the Markovian SIR process. The derivative of $y(t)$ in \eqref{eq_solution_SIR_y} equals
\begin{equation}\label{eq_solution_SIR_y_diff}
	y'(t) = \sum_{i=0}^{3^N-2^N-1} \tilde{c}_i \lambda_i e^{\lambda_i t}.
\end{equation}
The epidemic peak time $t_{\text{peak}}$ obeys $y'(t_{\text{peak}})=0$. We determine the epidemic peak time $t_{\text{peak}}$ using the Newton-Raphson and second-order Newton-Raphson method (see Appendix~\ref{app_root_finding} for the derivation). Methods like Newton-Raphson are suitable for the determination of the epidemic peak time using our solution, since all derivatives of the prevalence $y(t)$ can be straightforwardly computed.

\begin{figure*}[!ht]
    \centering
    \subfloat[\label{fig_SIR_NR1}]{%
        \includegraphics[width=0.46\textwidth]{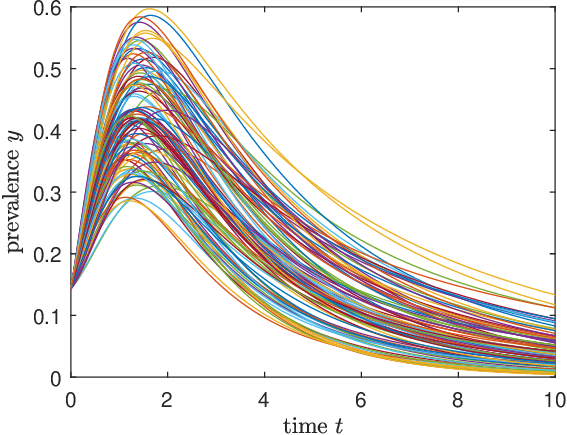}
    } \hfill
    \subfloat[\label{fig_SIR_NR2}]{%
        \includegraphics[width=0.46\textwidth]{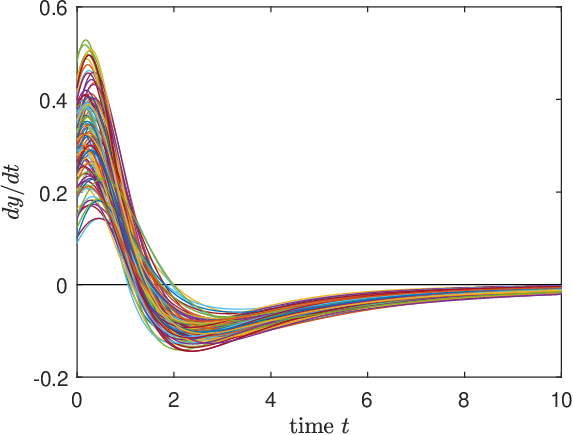}
    } \\
    \subfloat[\label{fig_SIR_NR3}]{%
        \includegraphics[width=0.46\textwidth]{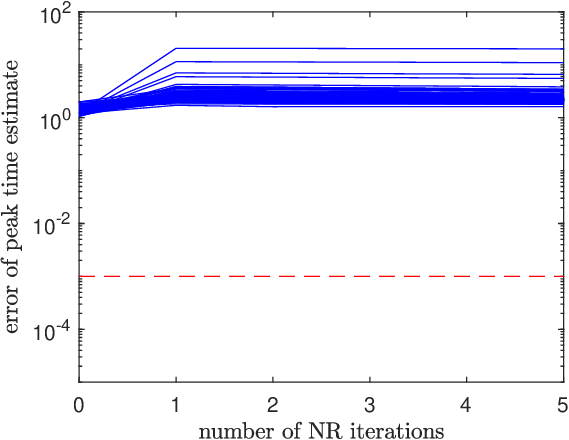}
    } \hfill
    \subfloat[\label{fig_SIR_NR4}]{%
        \includegraphics[width=0.46\textwidth]{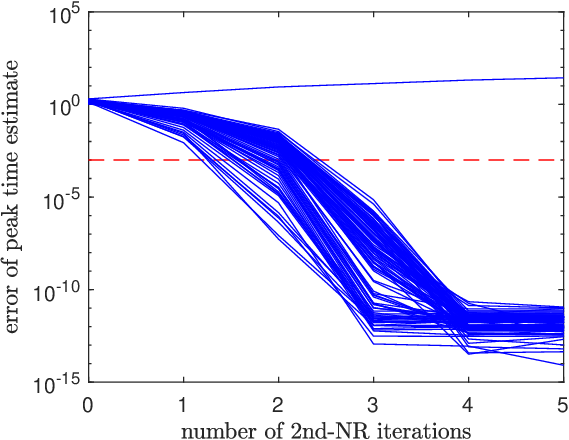}
    }
    \caption{(a) The time-varying prevalence $y(t)$ in SIR epidemics on the complete graph with $N=7$ nodes for 100 trials. Each trial randomly generated uniform random variables $\tilde{\beta}_{kl} \in [0.05,0.85]$ and $\delta_k \in [0.1,0.9]$. (b) The derivative of the prevalence $y$ with respect to time $t$ for each of the 100 trials. (c-d) The error of finding the true epidemic peak using the (c) Newton-Raphson method and the (d) second-order Newton-Raphson method with 0--5 iterations. Initially, $\tilde{t}_0=0$. The minimal error for finding the peak time is $10^{-10}$, because the \quotes{exact} peak is also determined numerically using the bisection method. Newton-Raphson in (c) does not converge, whereas second-order Newton-Raphson in (d) almost always converges quickly.}
    \label{fig_SIR_newton_raphson}
\end{figure*}

Starting with initial guess $\tilde{t}_0$, the Newton-Raphson method finds a root $t^*$ of $y'(t)$ iteratively as
\begin{equation}\label{eq_NR}
	\tilde{t}_{k+1} = \tilde{t}_k - \frac{y'(\tilde{t}_k)}{y''(\tilde{t}_k)},
\end{equation}
and converges, provided that the initial point $\tilde{t}_0$ is sufficiently close, to the root $t^*$. Performing Newton-Raphson on Eq.~\eqref{eq_solution_SIR_y_diff} with starting point $\tilde{t}_0=0$ yields for $k \geq 0$:
\begin{equation*}
	\tilde{t}_{k+1} = \tilde{t}_k - \frac{\displaystyle\sum_{i=0}^{3^N-2^N-1} \tilde{c}_i \lambda_i e^{\lambda_i \tilde{t}_{k}} }{\displaystyle\sum_{i=0}^{3^N-2^N-1} \tilde{c}_{i} \lambda_i^2 e^{\lambda_i \tilde{t}_{k}} }.
\end{equation*}
The Newton-Raphson method converges quadratically\footnote{Quadratic convergence means that the number of correct decimal digits roughly doubles with each iteration.} to a local root, which is not necessarily equal to a global maximum or minimum. Similarly, the second-order Newton-Raphson method is given by (see Appendix~\ref{app_root_finding})
\begin{equation}\label{eq_NR2}
    \tilde{t}_{k+1} = \tilde{t}_k + \frac{-y''(\tilde{t}_k) \pm \sqrt{y''(\tilde{t}_k)^2 - 2 y'(\tilde{t}_k) y'''(\tilde{t}_k)}}{y'''(\tilde{t}_k)}.
\end{equation}
We verify the accuracy of both approaches by comparing the peak time estimate with the true epidemic peak time $t_{\text{peak}}$ as follows: On the complete graph with $N=7$ nodes, we repeat 100 times: Draw uniformly distributed $\tilde{\beta}_{kl} \in [0.05,0.25]$ and $\delta_k\in[0.3,0.5]$ and compute for each set of parameters the exact peak time using Eq.\ \eqref{eq_solution_SIR_y_diff} and the bisection method. Figure~\ref{fig_SIR_NR1} shows the time-varying prevalence $y(t)$ for each of the 100 trials. The epidemic peak is approximately located at $t_{\text{peak}} \approx 2$. Figure~\ref{fig_SIR_NR3} shows the absolute difference between the real peak time $t_{\text{peak}}$ and the peak time estimated by the Newton-Raphson method for $k=0,\ldots,5$ iterations. For almost all trials, unfortunately, the error did not converge to zero. Contrary, the error of the second-order Newton-Raphson (2nd-NR) method in Figure~\ref{fig_SIR_NR4} rapidly converges to zero for almost\footnote{ It might happen that divergence occurs if the third derivative $y'''(t) \approx 0$, in which case a higher-order Newton-Raphson method should be implemented.} all trials. Due to the cubic convergence of 2nd-NR, only three iterations suffice to accurately approximate the true epidemic peak time $t_{\text{peak}}$.

\begin{figure}[!ht]
    \centering
    \includegraphics[width=0.48\textwidth]{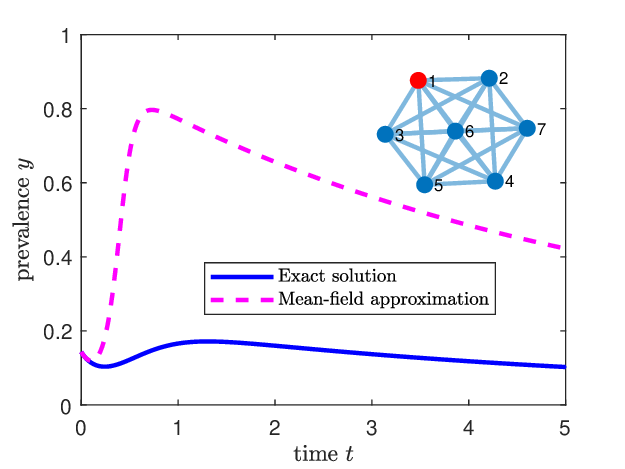}
    \caption{The exact solution (solid line) and mean-field approximation (dashed line) for SIR epidemics on the complete graph with $N=7$ nodes. The heterogeneous mean-field approximation from Eq.~(\ref{eq_SIR_MF}) poorly matches the exact solution. Additionally in this example, the Markovian prevalence $y(t)$ is not monotonic.} % (b) The derivative of the prevalence $y$ with respect to time.}
    \label{fig_example_SIR2}
\end{figure}

There are multiple reasons why the Newton-Raphson method does not converge to the true epidemic peak time $t_{\text{peak}}$. For example, for trials that satisfy $y'(0) > 0$ and $y''(0) > 0$ and starting at $\tilde{t}_0=0$, Eq.\ \eqref{eq_NR} tells us that $\tilde{t}_1 < 0$. Subsequent estimates $\tilde{t}_k$ for $k \geq 1$ will converge to even smaller (negative) values. For other trials, it holds that $y''(0)>0$ but $y''(0)$ is very small. The Newton-Raphson estimate $\tilde{t}_1$ is then so large, that subsequent estimates $\tilde{t}_k$ will diverge to $+\infty$, as $t^*=\infty$ is also a valid solution of $y'(t^*)=0$. Additionally, the prevalence $y(t)$ can be non-monotonic and exhibits multiple peaks. Figure~\ref{fig_example_SIR2} shows that such an effect can already appear in small graphs (although it is not very pronounced).

%We focus here on the simplified, but frequently occurring case with a single peak.

Figure~\ref{fig_example_SIR2} additionally demonstrates that the mean-field approximation performs very poorly. It is known that mean-field approximations are poor for small contact graphs \cite{ComparisonNIMFAandMarkov}, but allowing for heterogeneous transition rates leads to even worse mean-field estimates.

\subsection{Probability of $k$ infected nodes}
\begin{figure}[!ht]
	\centering
	\includegraphics[width=0.48\textwidth]{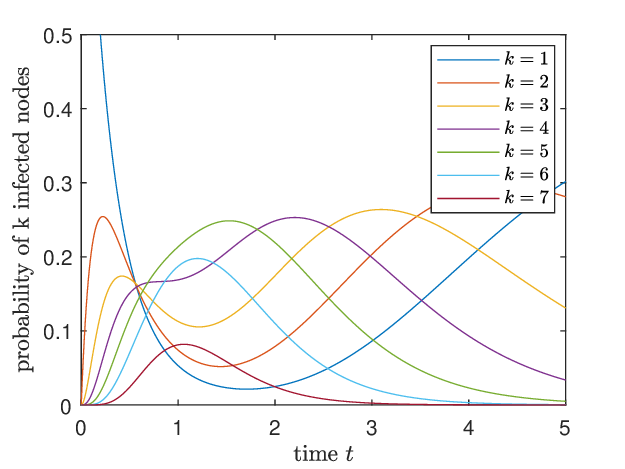}
	\caption{The probability of $k$ infected nodes in SIR epidemics on a complete graph with $N=7$ nodes and uniformly generated $\tilde{\beta}_{kl} \in [0.4, 0.9]$ and $\delta_k \in [0.1, 0.6]$. The probability of $k=1$ infected node equals 1 at $t=0$, whereas all other curves start at 0.}
	\label{fig_SIR_prob_k_infected}
\end{figure}

One of the advantages of our trinary numeral system for constructing the state space $\mathbf{S}$ of the $3^N$-sized Markov chain, is that many properties of the SIR process can be computed with ease. For example, the probability that $k$ nodes are simultaneously infected at time $t$ characterises an epidemic outbreak and can be used to assess the maximum impact of the disease on the population. Given the solution $\s(t)$, the probability of $k$ infected nodes can be computed by projecting on the vector $\mathbf{m}$, whose elements $m_i=1$ if configuration $i$ has $k$ infected nodes. In practice, we count the number of ones in $\mathbf{x}_i$ to compute $m_i$. Figure~\ref{fig_SIR_prob_k_infected} shows the probability of $k$ infected nodes in a contact graph with 7 nodes. At time $t=0$, the probability of 1 infected node is 1, whereafter it converges quickly to smaller values. In this example, the prevalence is rather high, implying that many nodes can be infected simultaneously. Figure~\ref{fig_SIR_prob_k_infected} illustrates the probability that all nodes are infected simultaneously is $0.1$ at its maximum value, which is very high.

\subsection{Probability of group-level infections}
\begin{figure}[!ht]
	\centering
	\includegraphics[width=0.48\textwidth]{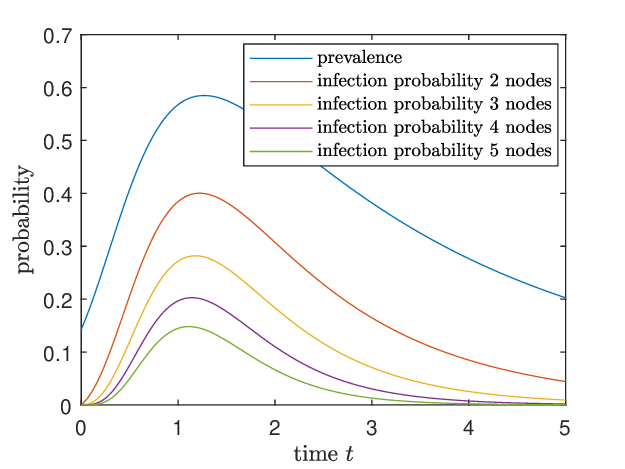}
	\caption{The joint infection probability of $k$ groups of nodes in SIR epidemics on a complete graph with $N=7$ nodes and uniformly generated $\tilde{\beta}_{kl} \in [0.4, 0.9]$ and $\delta_k \in [0.1, 0.6]$.}
	\label{fig_SIR_joint_prob}
\end{figure}

Many stochastic and deterministic epidemic models on networks focus on the prevalence or the probability of $k$ simultaneously infected nodes. The benefit of our exact method is the ability to exactly determine the probability that any group of $k$ nodes is simultaneously infected:
\begin{equation*}
	y^{(k)}(t) = \frac{1}{\binom{N}{k}} \sum_{\mathcal{S} \subseteq \mathcal{N}, |\mathcal{S}| = k} \P[X_{i_1} = 1, X_{i_2} = 1, \ldots, X_{i_k} = 1].
\end{equation*}
For $k=1$, we recover the prevalence $y(t)$. We know that $y^{(1)}(t) \geq y^{(2)}(t) \geq \ldots \geq y^{(N)}(t)$, because $\P[X=1,Y=1] \leq \P[X=1]$ for any two random variables $X$ and $Y$. Figure~\ref{fig_SIR_joint_prob} shows the joint infection probability of $k$ groups of nodes on a contact graph with $N=7$ nodes. All curves roughly exhibit their peak at the same epidemic peak time, but the height of the peak decreases with increasing $k$ as expected.

%We define the $3^N \times 1$ vector $\mathbf{m}^{(k)}$ to describe all configurations that contain $k$ simultaneously infected nodes. The solution follows as
%\begin{equation*}
%	y^{(k)} = \frac{1}{\binom{N}{k}} \sum_{i=0}^{3^N-1} \tilde{c}_i \left( \mathbf{m}^{(k)} \right)^T \w_i.
%\end{equation*}

The SIR epidemic model exhibits a phase transition around the epidemic threshold. Below the epidemic threshold, the disease dies out exponentially fast and above the threshold, the disease persists and infects a significant part of the population. It is conjectured that around the epidemic threshold, the probability of $k$ groups being infected is roughly equal for several values of~$k$. The epidemic threshold $\tau_c$ for Markovian SIR dynamics is not known, but lower-bounded by the mean-field threshold $\tau_c^{(1)}$, which is given in Eq.\ \eqref{eq_R0} in Appendix~\ref{app_mean_field}. Figure~\ref{fig_SIR_prob_group_infection} shows the peak time in (a) and height of the peak in (b). Around the epidemic threshold, which is approximated located at the normalised effective infection rate $x = \tau/\tau_c^{(1)} = 1$, we do not observe that all curves are approximately of the same order of magnitude. Most likely, a contact graph with $N=7$ nodes is too small to exhibit a clear epidemic threshold.

\begin{figure*}[!ht]
	\centering
	\subfloat[\label{fig_SIR_prob_group_infection_time}]{%
		\includegraphics[width=0.48\textwidth]{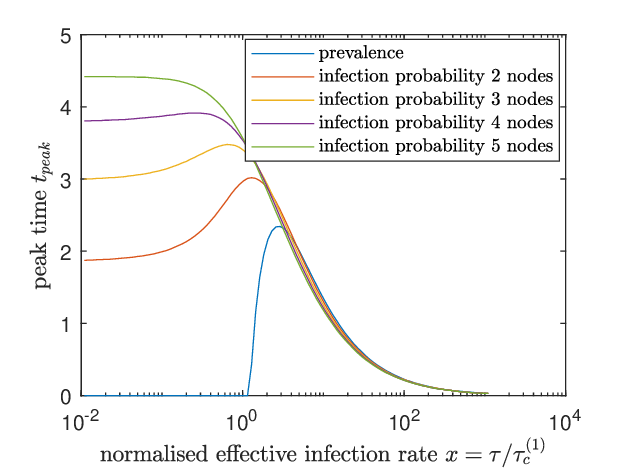}
	}
	\subfloat[\label{fig_SIR_prob_group_infection_prob}]{%
		\includegraphics[width=0.48\textwidth]{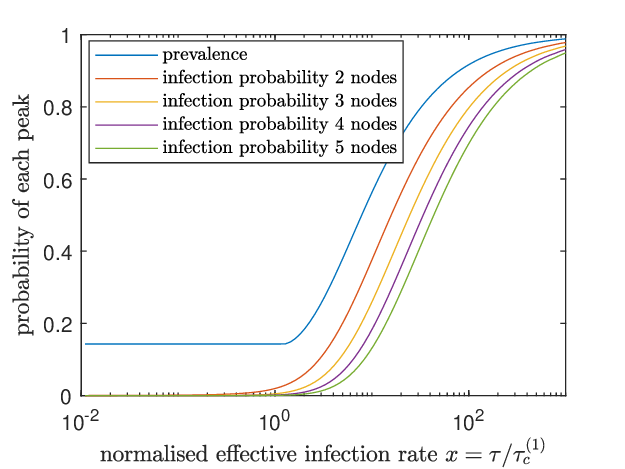}
	}
	\caption{Finding the (a) time and (b) size of the epidemic peak for $k$ groups of infected nodes in SIR epidemics on a complete graph with $N=7$ nodes and uniformly generated $\tilde{\beta}_{kl} \in [0.4, 0.9]$ and $\delta_k \in [0.1, 0.6]$. Around the $x=1$, which lower-bounds the epidemic threshold, all curves do not have the same magnitude.}
	\label{fig_SIR_prob_group_infection}
\end{figure*}

\subsection{Final epidemic size}
\begin{figure}[!ht]
    \centering
    \includegraphics[width=0.48\textwidth]{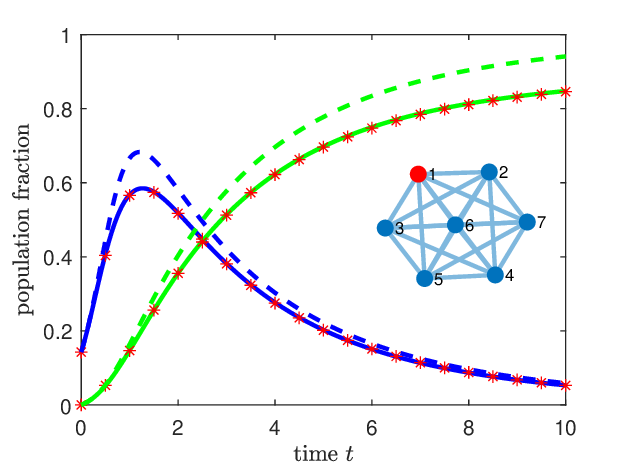}
    \caption{The average fraction of infected nodes (blue) and average fraction of recovered nodes (green) in a complete, weighted graph with $N=7$ nodes for the Markovian solution (solid), mean-field solution (dashed) and Monte Carlo simulations (asterisks). Parameters are the same as in Figure~\ref{fig_example_SIR}.}
    \label{fig_SIR_recovered}
\end{figure}

Of primary interest is the determination of the total number of infected nodes over the course of the disease outbreak, which is also known as the final epidemic size. The final epidemic size is equivalent to the average final fraction of recovered nodes. We split up the $3^N$ states in Eq.\ \eqref{eq_solution_SIR} into the set of $2^N$ absorbing states with eigenvalue $\lambda_i=0$ and the remaining states;
\begin{equation*}
    \s(t) = \sum_{i=0}^{2^N-1} c_i \w_i + \sum_{i=2^N}^{3^N-1} c_i e^{\lambda_i t} \w_i.
\end{equation*}
In the limit $t\to\infty$, the second term becomes zero. Then the steady-state distribution $\s_\infty$ is given by
\begin{equation*}
    \s_\infty = \sum_{i=0}^{2^N-1} c_i \w_i.
\end{equation*}
Denote the vector $\mathbf{m}_R = [0, 0, 1, 0, 0, 1, 1, 1, 2, 0, ...]^T$, whose elements $(m_R)_i$ describe the number of recovered nodes in configuration $i$. Then the final fraction $R_\infty$ of recovered nodes equals
\begin{equation}\label{eq_R_infty}
    R_\infty = \sum_{i=0}^{2^N-1} c_i \mathbf{m}_R^T \w_i.
\end{equation}
Equation~\eqref{eq_R_infty} is a closed-form solution for the final size of the epidemic. The mean-field approximation from Appendix~\ref{app_mean_field} does not allow for a closed-form solution, due to the non-linearity of the mean-field approximation. Figure~\ref{fig_SIR_recovered} shows the average fraction of infected and recovered nodes over time. Although the discrepancy between the curves looks small, the final epidemic size $R_\infty \approx 0.9$, whereas the mean-field estimates $R_\infty^{(1)} \approx 0.9999$, which is far off.

As a side note, we have chosen to compute the \emph{average} final size, but the final size \emph{distribution} can be obtained in a similar manner, by constructing a matrix $M_R$, whose element $(M_R)_{i,j}=1$ if in configuration $i$ contains exactly $j$ recovered nodes.

\subsection{Exceeding hospital capacity}
Hospitals have a limited capacity to treat patients simultaneously. Thus, the number of simultaneously infected people must not exceed the hospital capacity. If the hospital capacity is forecasted to be insufficient, counter-measures must be taken, e.g.\ taking social distancing measures or scaling up the hospital capacity. Given the disease parameters and the hospital capacity $c_h$, we can compute the probability of exceeding the hospital capacity as
\begin{equation*}
    p_h = \max_t \P[ I(t) > c_h] = \max_t \mathbf{m}_H^T \s(t),
\end{equation*}
where the element $(m_H)_i$ of the $3^N \times 1$ vector $\mathbf{m}_H$ describes if configuration $i$ contains more than $c_h$ infected nodes and $I(t)$ is the random variable describing the number of simultaneously infected nodes. Using \eqref{eq_solution_SIR}, we obtain
\begin{equation}\label{eq_hospital_capacity}
    p_h = \max_t \sum_{i=0}^{3^N-1} c_i e^{\lambda_i t} \mathbf{m}_H^T \w_i.
\end{equation}
If the probability to exceed the hospital capacity $p_h$ is too large, policymakers may suggest to close facilities, shops or schools. Using estimates of the impact of those closings on the infection rate allows for the computation of the new exceedance probability $p_h$. Since the exceedance probability is a rare event, standard Monte Carlo simulations are unsuitable for the determination of this probability. For example, if it is required that the hospital capacity is not exceeded with probability $p_h = 10^{-6}$, at least $10^7$ Monte Carlo simulations are required to accurately estimate the probability, demonstrating the necessity of our exact solution.

\subsection{The Laplace transform}
\begin{figure*}[!ht]
	\centering
	\subfloat[\label{fig_SIR_laplace1}]{%
		\includegraphics[width=0.48\textwidth]{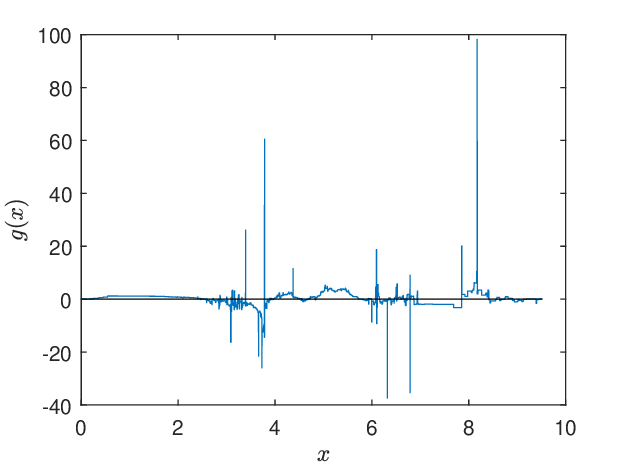}
	}
	\subfloat[\label{fig_SIR_laplace2}]{%
		\includegraphics[width=0.48\textwidth]{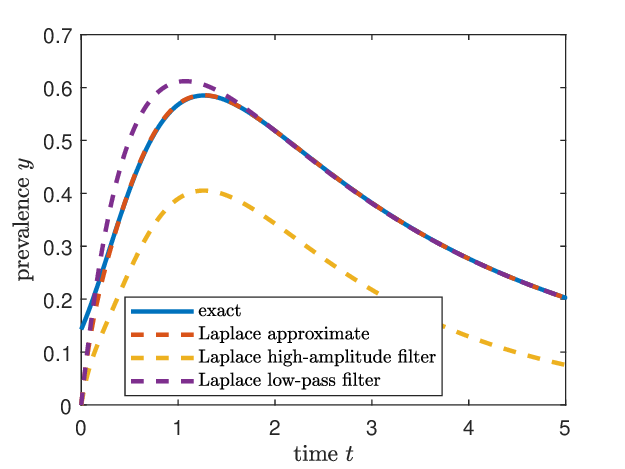}
	}
	\caption{(a) The function $g(x)$ from Eq.\ \eqref{eq_SIR_laplace} and (b) the prevalence \eqref{eq_SIR_laplace_prev} of the SIR process on a contact graph with $N=7$ nodes and uniformly generated $\tilde{\beta}_{kl} \in [0.4, 0.9]$ and $\delta_k \in [0.1, 0.6]$. Additionally, (b) shows the high-amplitude filter with cut-off amplitude $\epsilon=1$ and low-pass filter with frequency $x \in [0,2]$. The low-pass filter uses approximately 330 out of 2059 non-zero eigenvalues.}
	\label{fig_SIR_laplace}
\end{figure*}

In the limit of large networks, the solution \eqref{eq_solution_SIR_y2} is composed of exponentially many terms and cannot be computed. Fortunately, we can rewrite \eqref{eq_solution_SIR_y2} using Abel summation in the limit $N\to\infty$ to find (see Appendix~\ref{sec_Abel_summation} for the derivation):
\begin{equation}\label{eq_SIR_laplace_prev}
	y(t) = t \int_0^\infty e^{-xt} g(x) dx\hspace{1cm}(t>0)
\end{equation}
where
\begin{equation}\label{eq_SIR_laplace}
	g(x) = \sum_{l=0}^{[\lambda^{-1}(x)]} \tilde{c}_l.
\end{equation}
Here, $[x]$ indicates the integer part of $x$ and $\lambda^{-1}(x)$ is the inverse eigenvalue function. The function $g(x)$ is the inverse Laplace transform of $\frac{y(t)}{t}$ and contains all information about the prevalence, but is expressed in the frequency $x$ domain, rather than the time $t$ domain. For the complete graph with $N=7$ nodes, Figure~\ref{fig_SIR_laplace1} shows that the contribution $g(x)$ of each frequency $x$ exhibits a complicated structure. While \eqref{eq_SIR_laplace_prev} was derived in the limit $N\to\infty$, Figure~\ref{fig_SIR_laplace2} shows that the prevalence $y(t)$ from \eqref{eq_SIR_laplace_prev} for a contact graph with only 7 nodes approximately agrees with the exact solution~\eqref{eq_solution_SIR_y}. Equation~\eqref{eq_SIR_laplace_prev} is derived under the assumption $N\to\infty$, which is accurate everywhere, except around $t=0$. For a finite graph, the initial condition $y(0)=1/N$ is a finite number, whereas for $N \to \infty$, the initial prevalence converges to zero, causing a discrepancy between the two solutions. Nevertheless, for most networks of a realistic size, Eq.~\eqref{eq_SIR_laplace_prev} seems to be an accurate approximation of the solution~\eqref{eq_solution_SIR_y}.

Instead of the whole function $g(x)$, partial information on $g(x)$ may be sufficient to reconstruct the time-varying prevalence $y(t)$. We apply two techniques from signal processing to $g(x)$. Method 1 performs a high-amplitude filter, keeping only frequencies with amplitude larger than some bound $\epsilon$ and all other contributions are set to zero;
\begin{equation*}
	g_1(x) = \begin{cases}
		g(x), &\qquad g(x) \geq \epsilon \\
		0, &\qquad \text{otherwise}
	\end{cases}
\end{equation*}
Second, we employ a low-pass filter, only keeping the contributions of frequencies in the interval $[0, f_0]$;
\begin{equation*}
	g_2(x) = \begin{cases}
		g(x), &\qquad x \leq f_0 \\
		0, &\qquad \text{otherwise}
	\end{cases}
\end{equation*}
The main reason behind $g_2(x)$ is that small (in modulus) frequencies correspond to  small (in modulus) eigenvalues, which have the largest contributions in the Laplace transform \eqref{eq_SIR_laplace_prev}. Figure~\ref{fig_SIR_laplace2} shows the result of applying both filters to $g(x)$. The low-pass filter is superior to the high-amplitude filter, because small values of $x$ have a more significant contribution in the Laplace transform \eqref{eq_SIR_laplace_prev}. Even for an amplitude cut-off $\epsilon=1$ in Figure~\ref{fig_SIR_laplace2}, which is relatively small, the low-pass filter better approximates the true solution.

\begin{figure}[!ht]
	\centering
	\includegraphics[width=0.48\textwidth]{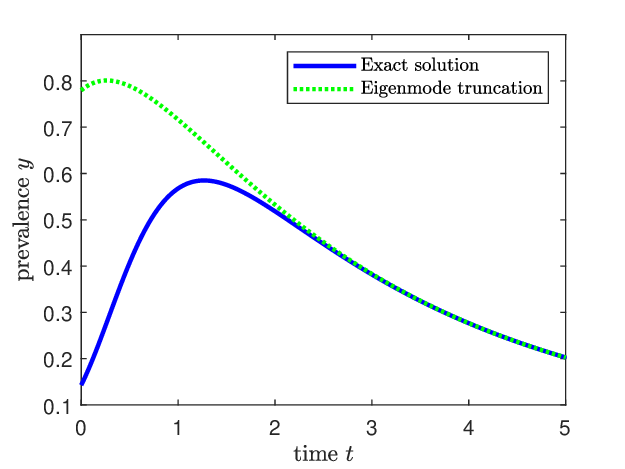}
	\caption{Eigenmode truncation in SIR epidemics on the complete graph with $N=7$ nodes and uniformly generated $\tilde{\beta}_{kl} \in [0.4, 0.9]$ and $\delta_k \in [0.1, 0.6]$. Approximately 330 out of 2059 non-zero eigenvalues are used.}
	\label{fig_SIR_eigenmode_truncation}
\end{figure}

Interestingly, eigenmode truncation (as explained in Section~\ref{sec_comp_infeasible}) is only effective for large times, as shown in Figure~\ref{fig_SIR_eigenmode_truncation}. Eigenmode truncation also maintains the smallest (in modulus) eigenvalues, but apparently performs much worse compared to the low-pass filter of the Laplace transform. The primary reason for this discrepancy is that eigenmode truncation considers finite networks whereas the Laplace transform was derived under the assumption of an infinitely large network. Even for a network with $N=7$ nodes, the Laplace transform (\ref{eq_SIR_laplace_prev}) performs better using the same number of eigenmodes. The advantage of the latter is that the solution can no longer explode around $t=0$ because the $t$-term in \eqref{eq_SIR_laplace_prev} guarantees $y(0)=0$.

\section{Conclusion}
\label{sec_conclusion}
We investigated continuous-time Markovian compartmental models on networks. We focussed on transient epidemic compartmental models such as SI and SIR, where re-infections do not appear. Then the compartmental graph describing the transitions between compartments does not contain directed cycles. We showed that such a continuous-time Markovian compartmental model admits an analytic solution using eigendecomposition of the infinitesimal generator. We demonstrated our method on the SI and SIR processes. We derived the eigenvalues of the infinitesimal generators and showed that the eigenvalues are related to S-I cut sets in the contact graph. Assuming that all non-zero eigenvalues are distinct, we constructed the exact solution of the SI and SIR process on heterogeneous networks. Additionally, we showed that the transient compartmental model can be extended by adding self-infections, simplicial contagion, temporal networks and non-Markovian dynamics while maintaining an exact, analytic solution for the time-varying infection probabilities of all nodes. 

The advantage of our exact approach is twofold. First, the analytic solution on small networks can serve as a calibration tool for (non)-Markovian compartmental simulators on networks and subsequently allows researchers to determine the number of simulation events to achieve a certain accuracy. Second, we showed that the underlying state space scales exponentially with the number of nodes. Much effort is devoted to simplify the state space, e.g.\ by aggregating states in the Markov graph. Our exact time-dependent solution can assess the quality of the proposed aggregation methods in an exact manner.

Finally, the advance in quantum computers and quantum algorithms may help to exactly determine the time-varying prevalence for much larger networks. There is an obvious analogy between the viral state of a node in a network and a quantum state of a particle in a quantum device. Also, the entire state space of a quantum system is increasing exponentially in the number of qubits. Hence, we expect a computational breakthrough and an unraveling of the SIR epidemic phase transition with the further development of quantum computers.

%We see three possible extensions of our results. So far, the exact solution can only be computed for heterogeneous networks. By relaxing the assumption of non-degenerate eigenvalues of the infinitesimal generator, the exact solution can be computed for homogeneous and undirected networks. Unfortunately, our exact method heavily relies on the uniqueness of eigenvalues, so we believe a completely different approach may be required for the homogeneous case. 
%
%A second research direction focusses on the exponentially large state space of the Markov chain. One can approximate the true solution by aggregating states in the exponentially large Markov graph. Using the exact time-dependent solution, the accuracy of the aggregation method can be studied, which will undoubtedly depend on the underlying graph.
%
%Finally, for practical applications, the infection parameters in the SI process are generally unknown and have to be estimated from measurements on the spreading process. Even though the Markov chain consists of exponentially many states, only the infection rate matrix with a quadratic number of terms have to be estimated. Furthermore, the SI process only consists of transient states, in which the return probability is zero. Thus we believe parameter estimation is possible for SI epidemics on networks, even for a small number of observations.

\textbf{Acknowledgements}
We thank Robin Persoons and Brian Chang for useful discussions on the infinitesimal generator for the GEMF framework.

This research has been funded by the European Research Council (ERC) under the European Union's Horizon 2020 research and innovation programme (grant agreement No 101019718).

\appendix

%{\Large\textbf{Appendix}}

\section{Newton's root-finding method}
\label{app_root_finding}
We shortly explain Newton's method of finding the root $t^*$ of a function $f(t)$. In our case, we are interested in the root of $y'(t)$ at $t_{\text{peak}}$, but we write $f(t)$ and $t^*$ for generality.

\subsection{First-order Newton-Raphson}
The Taylor expansion of the function $f(t^*)$ around $t^*=t$ equals
\begin{equation}\label{eq_Taylor_expansion}
\begin{split}
    f(t^*) &= f_0(t) + (t^*-t) f_1(t) \\
    &\quad+ (t^*-t) f_2(t) + \O((t^*-t)^3)
\end{split}
\end{equation}
where we defined
\begin{equation*}
    f_k(t) = \frac{f^{(k)}(t)}{k!}.
\end{equation*}
Since $t^*$ is a zero of $f(t)$, we find $f(t^*)=0$. Using only terms up to first order in \eqref{eq_Taylor_expansion}, we find
\begin{equation*}
	0 = f_0(t) + (t^*-t) f_1(t) + \O((t^*-t)^2)
\end{equation*}
which can be rewritten as
\begin{equation*}
	t^* = t - \frac{f_0(t)}{f_1(t)} + \O((t-t^*)^2)
\end{equation*}
Taking $t^* = \tilde{t}_{k+1}, t = \tilde{t}_k$ and neglecting the second-order terms, we find the Newton-Raphson iterative scheme:
\begin{equation*}
	\tilde{t}_{k+1} = \tilde{t}_k - \frac{f_0(\tilde{t}_k)}{f_1(\tilde{t}_{k})}.
\end{equation*}

\subsection{Second-order Newton-Raphson}
Instead of using only first-order terms, we can also use second-order terms in \eqref{eq_Taylor_expansion}. Again using $f(t^*)=0$, we find
\begin{equation*}
	0 = f(t) + (t^*-t) f_1(t) + (t^*-t)^2 f_2(t) + \O((t^*-t)^3)
\end{equation*}
Neglecting the third-order terms, we can solve for $t^*$;
\begin{equation*}
	t^* = t + \frac{-f_1(t) \pm \sqrt{f_1(t)^2 - 4 f_0(t) f_2(t)}}{2f_2(t)}
\end{equation*}
Using $t^*=\tilde{t}_{k+1}$ and $t=\tilde{t}_k$, we find the second-order Newton-Raphson iterative scheme:
\begin{equation*}
	\tilde{t}_{k+1} = \tilde{t}_k + \frac{-f_1(\tilde{t}_k) \pm \sqrt{f_1(\tilde{t}_k)^2 - 4 f_0(\tilde{t}_k) f_2(\tilde{t}_k)}}{2f_2(\tilde{t}_k)}
\end{equation*}
Substituting $f(t) = y'(t)$ gives Eq.\ \eqref{eq_NR2}. The major advantage of the second-order Newton-Raphson method is that the radius of convergence seems larger than Newton-Raphson and the speed of convergence is faster (cubic versus quadratic convergence).

%\subsection{Householder's method}
%One can also use the complete Taylor expansion \eqref{eq_Taylor_expansion} to solve for $t^*$ without %any iterative steps. The major issue is that the radius of convergence is extremely small...

%We find
%\begin{equation}
%    t_{k+1} = t_{k} - \frac{f_0(t_k)}{f_1(t_k)} - \frac{f_2(t_k)}{f_1(t_k)} \left( \frac{f_0(t_k)} %{f_1(t_k)} \right)^2 + \ldots
%\end{equation}

%
%\subsection{Picard}
%Picard iterations are simply described by
%\begin{equation}
%    t_{k+1} = t_k + f(t_k)
%\end{equation}
%which have a linear convergence, but a large radius of convergence.

\section{Mean-field equations}\label{app_mean_field}
The heterogeneous mean-field approximation of the stochastic SI(S) process is known as the N-Intertwined Mean-Field Approximation (NIMFA) \cite{IntroNIMFA} and equals
\begin{equation*}
	\df{v_k}{t} = - \delta_k v_k + (1-v_k) \sum_{j=1}^N \tilde{\beta}_{kl} v_l
\end{equation*}
where $v_k(t)$ is the probability that node $k$ is infected at time $t$ and $\tilde{\beta}_{kl}= \beta_{kl} a_{kl}$.

Similarly, the heterogeneous individual mean-field approximation for the SIR process is given by \cite{youssef2011individual}
\begin{align}
\begin{split}\label{eq_SIR_MF}
    \df{s_k}{t} &= - \sum_{l=1}^N \tilde{\beta}_{kl} s_k v_l \\
    \df{v_k}{t} &= \sum_{l=1}^N \tilde{\beta}_{kl} s_k v_l - \delta_k v_k
\end{split}
\end{align}
where $s_k(t)$ and $v_k(t)$ represent the probabilities that node $k$ is susceptible or infected at time $t$, respectively.

The mean-field SIS and SIR threshold is \cite{VandenDriesscheWatmough}
\begin{equation}\label{eq_R0}
	\tau_c^{(1)} = \lambda_{\max}(S^{-1} B)
\end{equation}
where $S = \text{diag}(\delta_1, \ldots, \delta_N )$ is the $N \times N$ curing rate matrix and $B$ is the $N \times N$ infection rate matrix, whose elements are $\tilde{\beta}_{kl}$.

\section{Abel summation}\label{sec_Abel_summation}
A finite-$n$-analysis on the sum $\sum_{k=m}%
^{n}a_{k}\,b_{k}$, where $n\geq m$, illustrates interesting formal
manipulations dealing with partial sums, $s_{k}=\sum_{l=m}^{k}a_{l}$ where
$k\geq m$, attributed to Niels Abel.

The basic observation is that $a_{k}=s_{k}-s_{k-1}$ for $k>m$. Thus,
\begin{align*}
	\sum_{k=m}^{n}a_{k}\,b_{k}  &  =a_{m}\,b_{m}+\sum_{k=m+1}^{n}(s_{k}%
	-s_{k-1})\,b_{k}\\
	&  =a_{m}\,b_{m}+\sum_{k=m+1}^{n}s_{k}\,b_{k}-\sum_{k=m}^{n-1}s_{k}\,b_{k+1}\\
	&  =a_{m}\,b_{m}+\sum_{k=m+1}^{n-1}s_{k}\,(b_{k}-b_{k+1}) \\
 &\qquad+s_{n}\,b_{n}%
	-s_{m}\,b_{m+1}%
\end{align*}
Since $s_{m}=a_{m}$, we arrive at Abel's partial summation valid for any
integer $m\leq n$,
\begin{equation}
	\sum_{k=m}^{n}a_{k}\,b_{k}=\sum_{k=m}^{n-1}\left(  \sum_{l=m}^{k}a_{l}\right)
	\,(b_{k}-b_{k+1})+b_{n}\left(  \sum_{l=m}^{n}a_{l}\right)
	\label{Abel_summation}%
\end{equation}
A reversal of the $k$- and $l$-sum again returns to the sum at left-hand side.

As an application of Abel summation, we consider the sum $h_{n}\left(
t\right)  =\sum_{k=0}^{n}a_{k}\,e^{-\lambda_{k}t}$, where $0\leq
\operatorname{Re}\left(  \lambda_{0}\right)  \leq\operatorname{Re}\left(
\lambda_{1}\right)  \leq\cdots\leq\operatorname{Re}\left(  \lambda_{n}\right)
$, that can appear as a solution of $n+1$-th order linear differential equation.
Abel summation (\ref{Abel_summation}) yields
\[
h_{n}\left(  t\right)  =\sum_{k=0}^{n-1}\left(  \sum_{l=0}^{k}a_{l}\right)
\,(e^{-\lambda_{k}t}-e^{-\lambda_{k+1}t})+e^{-\lambda_{n}t}\left(  \sum
_{l=0}^{n}a_{l}\right)
\]
Invoking $-t\int_{\lambda_{k+1}}^{\lambda_{k}}e^{-xt}dx=e^{-\lambda_{k}%
	t}-e^{-\lambda_{k+1}t}$, we obtain%
\[
h_{n}\left(  t\right)  =t\sum_{k=0}^{n-1}\int_{\lambda_{k}}^{\lambda_{k+1}%
}\left(  \sum_{l=0}^{k}a_{l}\right)  e^{-xt}dx+e^{-\lambda_{n}t}\left(
\sum_{l=0}^{n}a_{l}\right)
\]
Let the function $y=\lambda\left(  x\right)  $ define the sequence $\left\{
\lambda_{k}\right\}  _{0\leq k\leq n}$ by $\lambda\left(  k\right)
=\lambda_{k}$ at integer values of $x$. The inverse function $x=\lambda
^{-1}\left(  y\right)  $ maps the index to $k=\lambda^{-1}\left(  \lambda
_{k}\right)  $. Since $k\leq\lambda^{-1}\left(  x\right)  <k+1$ for
$\lambda_{k}\leq x<\lambda_{k+1}$, the integer part $\left[  .\right]  $ of
$\left[  \lambda^{-1}\left(  x\right)  \right]  =k$ for $\lambda_{k}\leq
x<\lambda_{k+1}$ and we have that%
\begin{align*}
	&\sum_{k=0}^{n-1}\int_{\lambda_{k}}^{\lambda_{k+1}}\left(  \sum_{l=0}^{k}%
a_{l}\right)  e^{-xt}dx \\
	&=\sum_{k=0}^{n-1}\int_{\lambda_{k}}^{\lambda_{k+1}%
}e^{-xt}\sum_{l=0}^{\left[  \lambda^{-1}\left(  x\right)  \right]  }%
a_{l}\;dx \\
&=\int_{\lambda_{0}}^{\lambda_{n}}e^{-xt}\sum_{l=0}^{\left[
	\lambda^{-1}\left(  x\right)  \right]  }a_{l}\;dx
\end{align*}
which leads to%
\begin{equation}
	h_{n}\left(  t\right)  =t\int_{\lambda_{0}}^{\lambda_{n}}e^{-xt}\sum
	_{l=0}^{\left[  \lambda^{-1}\left(  x\right)  \right]  }a_{l}\;dx+e^{-\lambda
		_{n}t}\left(  \sum_{l=0}^{n}a_{l}\right)  \label{Finite_Laplace_transform}%
\end{equation}
Relation (\ref{Finite_Laplace_transform}) shows for $t=0$ that $h_{n}\left(
0\right)  =\sum_{l=0}^{n}a_{l}$ for all $n$ (also when $n\rightarrow\infty$).

However, if $\lambda_{n}\rightarrow\infty$ for $n\rightarrow\infty$ and
$\lambda_{0}=0$, then the last sum in (\ref{Finite_Laplace_transform})
vanishes for all positive $t>0$ and
\begin{equation}
	\frac{h_{\infty}\left(  t\right)  }{t}=\frac{1}{t}\sum_{k=0}^{\infty}%
	a_{k}\,e^{-\lambda_{k}t}=\int_{0}^{\infty}e^{-xt}g\left(  x\right)
	dx\hspace{1cm}\left(  t>0\right)  \label{Laplace_transform}%
\end{equation}
is the Laplace transform of the sum $g\left(  x\right)  =\sum_{l=0}^{\left[
	\lambda^{-1}\left(  x\right)  \right]  }a_{l}$. When $t$ tends to zero as $t=\frac{c}{\lambda_n}$ in the limit $n\rightarrow\infty$, then  $e^{-\lambda
		_{n}t}=e^{-c}$, which is a constant,  and the first term in
(\ref{Finite_Laplace_transform}) vanishes, but the second term remains and $e^{-c}=1$ so that
$h_{\infty}\left(  0\right)  =\sum_{l=0}^{\infty}a_{l}$, which is a prerequisite for continuity for $t\geq 0$ as well. Hence, the point
$t=0$ needs care\footnote{If $h_{\infty}\left(  0\right)  =0$, then the limit
	$\lim_{t\rightarrow0}\frac{h_{\infty}\left(  t\right)  }{t}=h_{\infty}%
	^{\prime}\left(  0\right)  $ and (\ref{Laplace_transform}) shows that
	$\int_{0}^{\infty}g\left(  x\right)  dx=h_{\infty}^{\prime}\left(  0\right)
	$.}.
 
 Observe from $g\left(  x\right)  =\sum_{l=0}^{\left[  \lambda^{-1}\left(
	x\right)  \right]  }a_{l}$ in the limit $n\rightarrow\infty$ where
$\lambda_{n}\rightarrow\infty$, that $g\left(  0\right)  =a_{0}=\lim
_{t\rightarrow\infty}h_{\infty}\left(  t\right)  $, while $\lim_{x\rightarrow
	\infty}g\left(  x\right)  =\sum_{l=0}^{\infty}a_{l}=h_{\infty}\left(
0\right)  $. The inverse Laplace transform is%
\[
g\left(  x\right)  =\sum_{l=0}^{\left[  \lambda^{-1}\left(  x\right)  \right]
}a_{l}=\frac{1}{2\pi i}\int_{c-i\infty}^{c+i\infty}\frac{h_{\infty}\left(
	t\right)  }{t}e^{xt}dt\hspace{1cm}\left(  c>0\right)
\]

The infinitesimal generator $Q$ is minus a weighted Laplacian matrix and the
smallest eigenvalue of any Laplacian matrix is zero. Hence, $\lambda_{0}=0$ is
satisfied, but $\lambda_{n}$ is always finite for any finite graph with $n$
nodes. Only if the graph is sufficiently large, then the Laplace transform
(\ref{Laplace_transform}) is exact for $t>0$, while $h_{n}\left(  0\right)
=\sum_{l=0}^{n}a_{l}$ for all $n$.

\section{General compartments}\label{sec_gemf}
\begin{figure*}[!htb]
	\centering
	\includegraphics{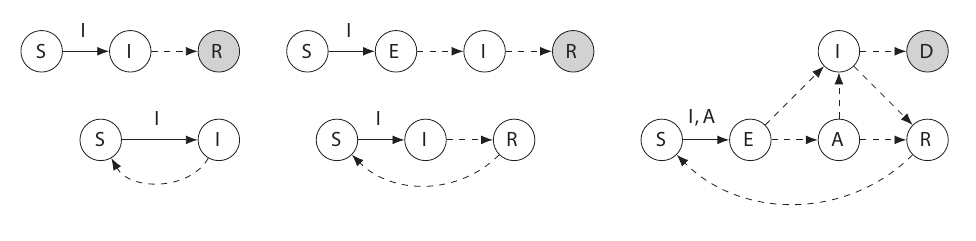}
	\caption{Several well-known compartmental epidemic models. Left: SIS and SIR. Middle: SEIR and SIRS. Right: SEIARDS. Absorbing states are shown as grey nodes. Full links described link-based transitions (and the letter(s) on the link describe the compartment(s) that triggers the transition) whereas dashed links indicate node-based transitions. Here, the exposed component E includes nodes that have already been infected but are not yet infectious. Furthermore, A represents asymptomatic nodes that do not show any symptoms but can still infect susceptible nodes (although at a lower rate) and D represents the deceased nodes.}
	\label{fig_compartmental_graphs}
\end{figure*}

The Generalised Epidemic Mean-Field (GEMF) framework describes general compartmental epidemic models on networks \cite{GEMF2}. Each node~$k$ at time $t$ is in one of the $c$ possible compartments. The compartmental graph describes all possible transitions between the different compartments and additionally specifies whether the transition is a link-based or a node-based transition. Figure~\ref{fig_compartmental_graphs} shows several examples of well-known compartmental graphs in epidemiology. As shown in Figure~\ref{fig_compartmental_graphs}, multi-links can exist in the compartmental graph if multiple processes cause the same transition. For example, infections may happen based on contact between infected and susceptible nodes, but susceptible nodes can also be infected indirectly, which is modelled as a self-infection process. Both the infection and self-infection process (in $\eps$-SIR) change the viral state of a node from susceptible to infected, thus allowing for multi-links in the compartmental graph.

We use a base-$c$ numeral system to denote each of the $c$ possible viral states of each individual. We denote $x_k(i)$ as the viral state of node $k$ in configuration $i$, which can be any integer between $0$ and $c-1$. The viral state vector $\x = (x_N, x_{N-1}, \ldots, x_1)^T$ represents the viral state of all nodes. Given a particular viral state vector $\x$, we compute the configuration number $i$ (which is a decimal number) based on the base-$c$ numerals as
\begin{equation}\label{eq_configuration_number}
	i = \sum_{k=1}^N x_k(i) \cdot c^{k-1}
\end{equation}
Representation \eqref{eq_configuration_number} guarantees that each viral state vector $\x$ receives a unique configuration number~$i$. The total number of configurations equals $c^N$, because each node has $c$ possible viral states and there are $N$ nodes in the contact graph.

The transitions in the Markov chain with $c^N$ states are described by the $c^N \times c^N$-dimensional infinitesimal generator $Q$. We assume that the transitions can be subdivided into a set $\LTset$ of link-based transitions and a set $\NTset$ of node-based transitions. Each link-based transition $\LTel$ describes the transition of node $l$ to change from compartment $\LTel_1$ to $\LTel_2$ with rate $\beta_{\LTel,kl} a_{kl}$ under the influence of an adjacent node $k$ whose current state is $\LTel_3$. Additionally, a node-based transition $\NTel$ changes node $l$ from state $\NTel_1$ to $\NTel_2$ with rate $\delta_{\NTel,l}$.

Due to the Markov property, any transition in the Markov graph will change the viral state of exactly one node. Thus the viral state vector $\x$ before and after the transition will differ by exactly one bit. The \emph{Hamming distance} $d_H(\x,\y)$ between two vectors $\x$ and $\y$ is the number of elements that is different in $\x$ and $\y$. Transitions in the Markov graph occur if a single node changes its state, which corresponds to states $\x$ and $\y$ having Hamming distance $d_H(\x,\y)=1$. Note that the reverse does not hold, i.e.\ having $d_H(\x,\y)=1$ does not necessarily imply that a transition is possible, because transitions are dependent on the existence of directed links in the compartmental graph (recall Figure~\ref{fig_compartmental_graphs}).

For general Markovian compartmental models with $c$ compartments, the set $\LTset$ of link-based transitions and the set $\NTset$ of node-based transitions, the infinitesimal generator $Q$ can be constructed as follows:
\begin{subequations}\label{eq_Q_gemf}
    \begin{align}
        \begin{split}
            q_{ij} &= \sum_{\LTel \in \LTset} \sum_{k=1}^N \beta_{\LTel,kl} a_{kl} \mathbf{1}_{\{x_l(i) = \LTel_1 \cap \, x_l(j) = \LTel_2 \cap \, x_k(i) = \LTel_3\}} \\
            &\qquad\qquad+ \sum_{\NTel \in \NTset} \delta_{\NTel,l} \mathbf{1}_{\{x_l(i) = \NTel_1 \cap \, x_l(j) = \NTel_2\}} \\
            &\qquad\text{if } d_H(\x(i), \x(j)) = 1 \text{ and } l = \lfloor \log_c(|i-j|) \rfloor + 1 \\
        \end{split} \\
        q_{ii} &= -\sum_{\substack{j=0 \\ j\neq i}}^{c^N-1} q_{ij}
    \end{align}
\end{subequations}
where $\log_c(\ldots)$ is the base-$c$ logarithm and $\lfloor x \rfloor$ is $x$ rounded down to the nearest integer. Transitions from configuration $i$ to $j$ only occur by changing the viral state of exactly one node, thus exactly one bit in the viral state vector $\x$. For a given configuration $i$, the maximum possible number of transitions is given by the maximum out-degree $d_{\max}$ of the compartmental graph (recall Figure~\ref{fig_compartmental_graphs}), because the out-degree of a viral state specifies the number of possible next viral states. For a given configuration $i$, the number of possible transitions is given by $d_{\max} N$, because each node can change its viral state. Taking into account the possibility of not making any change, the number of non-zero elements in the infinitesimal generator $Q$ is upper bounded by $(d_{\max} N + 1) c^N$, which is significantly less than the dense $c^N \times c^N$ matrix.

Using the infinitesimal generator $Q$ for any GEMF epidemic model, one can perform the same analysis as in the main body in this paper, i.e.\ analysing the eigenvalues and eigenvectors and computing the exact time-dependent solution.

\section{Paths in the SIR Markov graph}\label{app_SIR_Markov_graph}

Each state in the SIR Markov graph has maximally $N$ out-going links, i.e.\ the maximum out-degree is $N$. The reason is that every state describes the viral state of $N$ nodes. During the Markovian SIR process, only one event can occur simultaneously, which results in one node changing its state. Since the graph consists of $N$ nodes, at most $N$ states can be reached from the starting state. Similarly, the maximum in-degree equals $N$.

%The \emph{number of nodes in each layer $l$} are equal to the $l$th element from the $N$th row of the Trinomial triangle, which is a variation of Pascal's triangle. Those values are also called the trinomial coefficients. For general $N$ and layer $l$, the number of nodes equals
%\begin{equation*}
%    \begin{pmatrix}
%        N \\ l
%    \end{pmatrix}_2 = \sum_{0 \leq a,b \leq N, a+2b=N+l} %\frac{N!}{a! \, b! \, (N-a-b)!}
%\end{equation*}

We are interested in the \emph{number of paths between state $i$ and $j$}. The maximum number of paths is attained if (i) state $i$ corresponds to the initial state, in which one node is infected and the other nodes are susceptible, (ii) state $j$ corresponds to the all-recovered state and (iii) the contact graph~$G$ equals the complete graph. Then we obtain the following result:
\begin{widetext}
    \begin{lemma}\label{lem_max_paths}
        The maximum number of paths $|\mathcal{P}|$ in the SIR Markov graph for the \textbf{complete graph} with $N \geq 3$ is bounded by
        \begin{equation}\label{eq_no_paths_bounds}
            (N-1)! \cdot N! < (2N-1) ((N-1)!)^2 < |\mathcal{P}| < \frac{(2N-1)!}{2^{N-1}} < \frac{(2N)!}{2^N} < N^{2N-1}.
        \end{equation}
    \end{lemma}
\end{widetext}
\textit{Proof}.\ \textbf{Upper bound 1:} The amount of paths between state $i$ and $j$ is maximally equal to $N^{2N-1}$, because the maximal hopcount equals $2N-1$ and the maximum out-degree equals $N$.  The upper bound is never attained, because many states have out-degree smaller than $N$.
	
\textbf{Upper bound 2:} Consider adding a positive self-infection rate $\eps > 0$ to the SIR process. Then the Markov graph simplifies to only one absorbing state: the all-recovered state. Not starting from the initial state with a single infected node, but from the all-healthy state, we definitely have an upper bound on the number of paths in the SIR process. The number of paths in this case is equivalent to asking in how many ways the $N$ nodes can change from S to I, and from I to R. In general, there are $(2N)!$ ways. However, for a given node, it is required that the node gets infected before the node recovers. In exactly half of the cases, the ordering is wrong, thus we obtain $(2N)!/2$ paths. Repeating this argument for all $N$ nodes leads to $(2N)!/(2^N)$ possible paths.
	
\textbf{Upper bound 3:} In our case, the process is initiated with a single infected node. Using the argument from upper bound 2 for the $N-1$ initially non-infected nodes, we find $(2N-2)!/(2^{N-1})$ paths. However, besides the transition from the $N-1$ initially susceptible nodes from $S \to I \to R$, the initially infected node must also recover. There are exactly $2N-1$ ways to place the recovery of the initial node in the sequence of the total $2N-1$ events. This leads to $(2N-1)!/(2^{N-1})$ paths. This number of paths is still an upper bound, because we allow for transitions from absorbing states to non-absorbing states, which is not possible in the SIR process.
		
\textbf{Lower bound 1:} Computing all paths is complicated, so we restrict ourselves here to a subset of all possible paths. We only consider paths that pass through the all-infected state. Starting from a single infected node, the other $N-1$ nodes must become infected and the only thing that matters is the order in which the nodes are infected. This equals to $(N-1)!$ possible paths. From the all-infected state to the all-recovered state, all nodes must recover, and the only thing that matters is the order in which the nodes recover. This yields another $N!$ possible paths. In total, this equals $(N-1)! \cdot N!$ paths that pass through the all-infected state. Naturally, there are other paths too, causing the lower bound to be non-tight. 
	
\textbf{Lower bound 2:} The other paths might pass through a state with all nodes infected, except one node that remains susceptible and another is already recovered (this only happens if $N \geq 3$). There are in total $(N-1)^2$ of such paths, because the susceptible node can be chosen among all that are not initially susceptible and the recovered node can be chosen among all nodes not remaining susceptible. We split up the computation in two parts:
	
Part 1: from initial state to $(111111102)$ state. \\
In total $N-2$ nodes must be infected and 1 node must recover. To prevent issues, we assume that the recovery event is the last event. This is definitely a lower bound. Then there are $(N-2)!$ possible paths.
	
Part 2: from $(1111111102)$ state to all-recovered state $(2222222222)$.\\
In total $1$ node must be infected and $N-1$ nodes must recover. To prevent issues, we assume that the infection event is the first event. Again, this is definitely a lower bound. Then there are $(N-1)!$ possible paths.

In total, we conclude that the number of paths through state of the form $(11111111102)$ is lower-bounded by $(N-1) ((N-1)!)^2$. Hence, the lower bound can be improved to $(2N-1) ((N-1)!)^2$ for $N \geq 3$. \hfill $\square$

\begin{corollary}\label{cor_max_paths}
    For any \textbf{connected contact graph} $G$, the maximum number of paths $|\mathcal{P}|$ in the SIR Markov graph is bounded by
    \begin{equation*}
        N! \leq |\mathcal{P}| < \frac{(2N-1)!}{2^{N-1}}
    \end{equation*}
\end{corollary}
\textit{Proof.} The upper bound follows from Lemma~\ref{lem_max_paths}. The lower bound follows the same reasoning as the first lower bound in Lemma~\ref{lem_max_paths}: we only consider the subset of paths that pass through the all-infected state. The number of paths from the initial state to the all-infected state is at least 1 in a connected graph, but the number of paths from the all-infected state to the all-recovered state equals $N!$, as the recovery of each node may happen independently of each other. \hfill $\square$

Corollary~\ref{cor_max_paths} implies that the maximum number of paths grows more than exponentially in the number of nodes~$N$. For the complete graph, Table~\ref{tab_paths} shows a numerical evaluation of the number of paths~$|\mathcal{P}|$, which is computed by enumerating all paths in the $3^N$-sized Markov graph. The combinatorial explosion of the number of paths prevents the computation of the exact SIR solution on any network of realistic size.

\begin{table}[!ht]
    \centering
    \caption{The maximal number of paths in the SIR Markov graph in the complete graph between the initial state with a single infected node and all other nodes susceptible, towards the all-recovered state on a complete contact graph~$G$. The lower bounds (LB) and upper bounds (UB) are from Eq.\ \eqref{eq_no_paths_bounds}.}
    \begin{tabular}{crrrrrr} %c|r|r|r|r|r|r
        \hline \hline
        $N$ & LB1 & LB2 & $|\mathcal{P}|$ & UB1 & UB2 & UB3 \\
        \hline 
        2 & 2 & - & 2 & 3 & 6 & 8 \\
        3 & 12 & 20 & 20 & 30 & 90 & 243 \\
        4 & 144 & 252 & 444 & 630 & 2,520 & 16,384 \\
        5 & 2,880 & 5,184 & 16,944 & 22,680 & 113,400 & 1,952,125 \\
        6 & 86,400 & 158,400 & 979,440 & 1,247,400 & 7,484,400 & $3.6 \cdot 10^8$ \\
        \hline\hline
    \end{tabular}
    \label{tab_paths}
\end{table}

\subsection{Recursive formula for the complete graph}
We compute the maximum number of paths in the complete graph exactly using a recursive formula. As before, the initial state is the state with one infected node and the other nodes are susceptible. The final state corresponds to the all-recovered state. Let $r_{kl}$ be the number of paths, starting from the initial state to a state that has $k$ infected nodes and $l$ recovered nodes, which is denoted by $(k,l)$. The number of paths $r_{kl}$ can be computed iteratively, by looking at the previous transition, which is either related to the infection or recovery of a node. Starting with $k-1$ infected nodes and $l$ recovered nodes, one of the susceptible nodes can get infected. Since we assume that the contact graph is the complete graph, all of the infected nodes can infect the susceptible nodes, leading to $N-k-l+1$ possible paths from state $(k-1,l)$ to $(k,l)$. Similarly, starting in the state with $k+1$ infected nodes and $l-1$ recovered nodes, we make a transition to $(k,l)$ if one of the infected nodes recovers. Since there are $k+1$ infected nodes, each of those nodes can recover, leading to $k+1$ possible paths from $(k+1,l-1)$ to $(k,l)$. Then the following recursion for $r_{kl}$ is obtained:
\begin{eqnarray}
    r_{kl} &=& (N-k-l+1) \, r_{k-1,l} + (k+1) \, r_{k+1,l-1} \label{SIRpathrecursion}\\
    r_{10} &=& 1. \qquad\qquad\qquad \textnormal{by definition}
   \nonumber
\end{eqnarray}
Since the process is initiated at $k=1$ infected node and $l=0$ recovered nodes, we define $r_{10}=1$. The number of paths from the initial state to the all-recovered state equals $r_{0N}$. %, while $r_{0l}=0$ for $l<N$. 

\begin{figure}[!htb]
    \centering
    \includegraphics[width=0.33\textwidth]{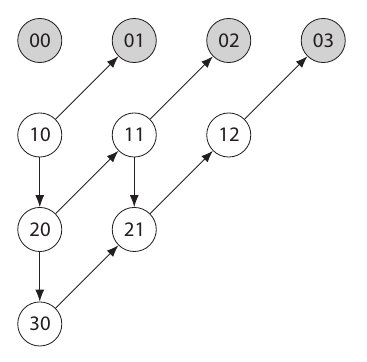}
    \caption{The simplified SIR Markov graph on a complete contact graph with $N$ nodes, here shown with $N=3$. The node labels $kl$ denote the number of infected nodes $k$ and the number of recovered nodes $l$.}
    \label{fig_Markovgraph_complete}
\end{figure}

Numerical evaluation of the recursion (\ref{SIRpathrecursion}) for $r_{kl}$ indicates that the number of paths equals
\begin{equation*}
    |\mathcal{P}| = (N-1)! \cdot f(N)
\end{equation*}
where $f(N)$ is given by \url{https://oeis.org/A000698}. Unfortunately, $f(N)$ does not seem to possess a closed-form solution.
Also, the generating function $R(x,y) = \sum_{k=0}^\infty \sum_{l=0}^\infty r_{kl} x^k y^l$, applied to the recursion (\ref{SIRpathrecursion}) and resulting in the partial differential equation
\begin{equation*}
     (x^2 - y) \pd{R}{x} + xy \pd{R}{y}-(Nx -1) \, R(x,y)=0
\end{equation*}
does not seem to admit a closed-form solution.

\begin{figure*}[!htb]
    \centering
    \subfloat[\label{fig_Markovgraph_SIR1} Compartmental graph]{%
        \includegraphics[width=0.3\textwidth]{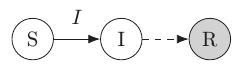}
    }
    \subfloat[\label{fig_Markovgraph_SIR2} Markov graph]{%
        \includegraphics[width=0.24\textwidth]{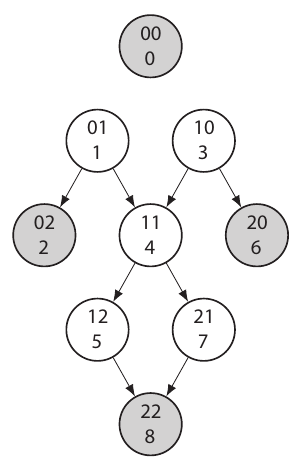}
    }
    \caption{The spread of SIR epidemics on a network with two connected nodes. (a) the SIR compartmental graph and (b) the SIR Markov graph with $3^2$ states.}
    \label{fig_Markovgraph_SIR}
\end{figure*}

\section{SIR exact solution for small graphs}\label{app_SIR_exact_solution}

The SIR equations can be solved explicitly for small graphs. Here, we provide those exact solutions on the complete and path graph with $N=2$ and $N=3$ nodes.

\subsection{Solution on $N=2$ nodes}
For $N=2$ nodes, the Markov graph contains 9 states and is shown in Figure~\ref{fig_Markovgraph_SIR}. Starting in the initial state $i=1$, that is, node~1 being infected and node~2 susceptible, the exact solution $\s(t)$ equals:
\begin{widetext}
\begin{align*}
    s_{0}(t) &= 0 \\
    s_{1}(t) &= e^{-(\beta_{21} + \delta_1)t} \\
    s_{2}(t) &= \frac{\delta_1}{\beta_{21} + \delta_1} \left(1 - e^{-(\beta_{21} + \delta_1)t} \right) \\
    s_{3}(t) &= 0 \\
    s_{4}(t) &= \frac{\beta_{21}}{\beta_{21} - \delta_2} \left( e^{-(\delta_1 + \delta_2)t} - e^{-(\beta_{21} + \delta_1)t} \right) \\
    s_{5}(t) &= \frac{\beta_{21}}{\beta_{21} + \delta_1 - \delta_2} e^{-\delta_2 t} + \frac{\beta_{21}}{\delta_2 - \beta_{21}} e^{-(\delta_1 + \delta_2)t} + \frac{\beta_{21} \delta_1}{(\beta_{21} - \delta_2)(\beta_{21} + \delta_1 - \delta_2)} e^{-(\beta_{21} + \delta_1)t} \\
    s_{6}(t) &= 0 \\
    s_{7}(t) &= e^{-\delta_1 t} + \frac{\beta_{21}}{\delta_2 - \beta_{21}} e^{-(\delta_1+\delta_2)t} + \frac{\delta_2}{\beta_{21}-\delta_2} e^{-(\beta_{21}+\delta_1)t} \\
    s_8(t) &= \frac{\beta_{21}}{\beta_{21} + \delta_1} - e^{-\delta_1 t} - \frac{\beta_{21}}{\beta_{21} + \delta_1 - \delta_2} e^{-\delta_2 t} + \frac{\beta_{21}}{\beta_{21} - \delta_2} e^{-(\delta_1 + \delta_2)t} \\
    &\qquad+ \left( \frac{\delta_1}{\beta_{21} + \delta_1} - \frac{\delta_2}{\beta_{21} - \delta_2} + \frac{\delta_2 - \delta_1}{\beta_{21} + \delta_1 - \delta_2} \right) e^{-(\beta_{21} + \delta_1)t}
\end{align*}
The prevalence follows by a weighted sum over the states $\mathbf{s}(t)$, where the weight $w_i$ equals the number of infected nodes in state $i$. Then we find
\begin{equation}
    y(t) = \frac{1}{2} \left( e^{-\delta_1 t} + \frac{\beta_{21}}{\beta_{21} + \delta_1 - \delta_2} e^{-\delta_2 t} - \frac{\beta_{21}}{\beta_{21}+\delta_1-\delta_2} e^{-(\beta_{21} + \delta_1) t} \right)
\end{equation}
and the average fraction of recovered nodes $r(t)$ equals
\begin{align}
    r(t) &= \frac{1}{2} \left( 1 + \frac{\beta_{21}}{\beta_{21} + \delta_1} - e^{-\delta_1 t} - \frac{\beta_{21}}{\beta_{21} + \delta_1 - \delta_2} e^{-\delta_2 t} + \frac{\beta_{21} \delta_2}{(\beta_{21} + \delta_1)(\beta_{21} + \delta_1 - \delta_2)} e^{-(\beta_{21} + \delta_1)t} \right)
\end{align}
Figure~\ref{fig_SIR_numerics} shows the solution of the exact SIR model for two cases: (left) below the epidemic threshold and (right) above the epidemic threshold. Below the threshold, the probability to arrive at the state $s_8$ is at most 0.33 whereas it is over 0.6 above the threshold.

\begin{figure*}[!ht]
    \centering
    \captionsetup[subfigure]{justification=centering}
    \subfloat[\label{fig_scenario1_s}]{%
        \includegraphics[width=0.49\textwidth]{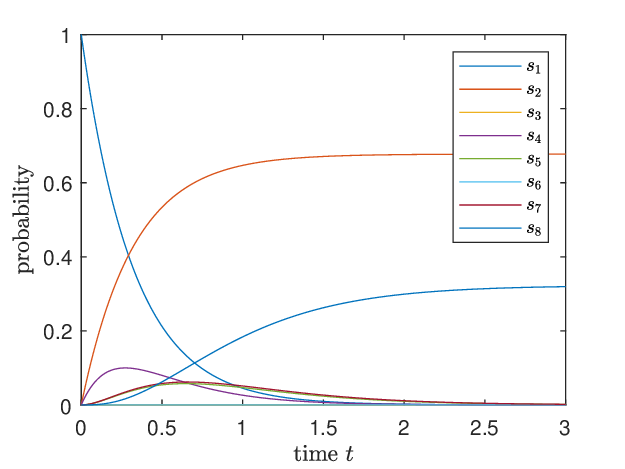}
    }
    \subfloat[\label{fig_scenario2_s}]{%
        \includegraphics[width=0.49\textwidth]{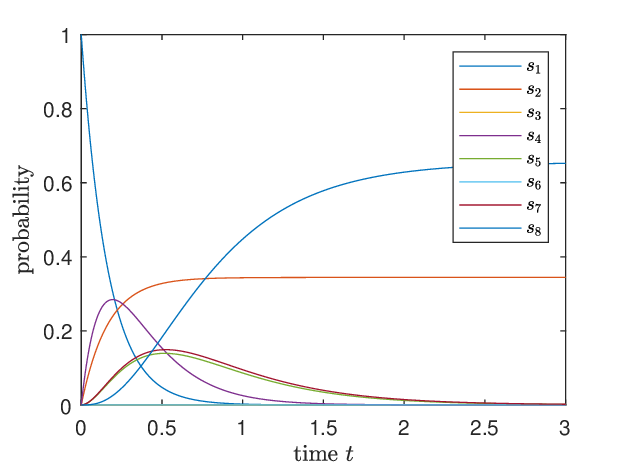}
    }\\
    \subfloat[\label{fig_scenario1_y}]{%
        \includegraphics[width=0.49\textwidth]{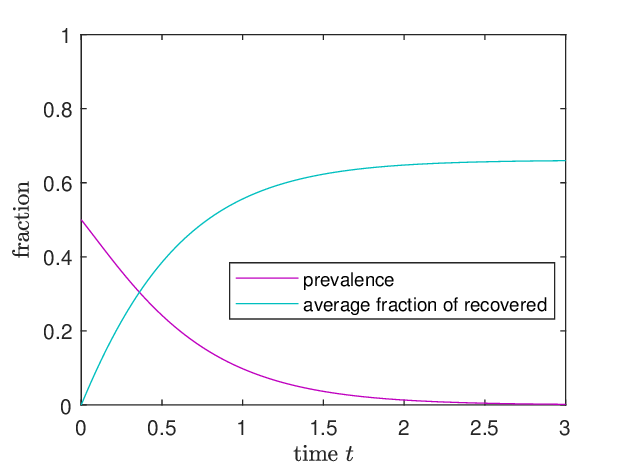}
    }
    \subfloat[\label{fig_scenario2_y}]{%
        \includegraphics[width=0.49\textwidth]{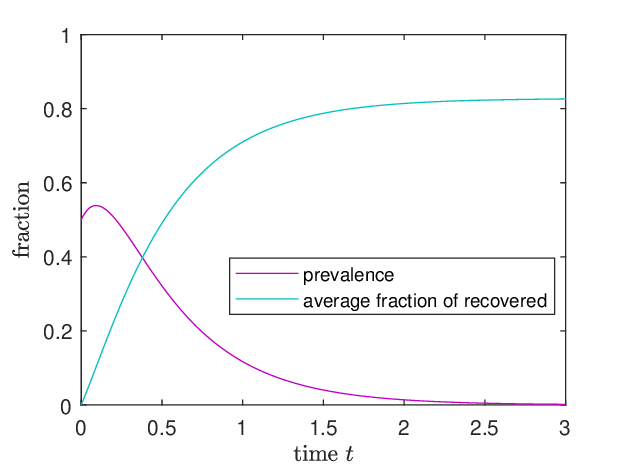}
    }
    \caption{Simulation of the SIR model on a graph with $N=2$ nodes. The recovery rates are $\delta_1 = 2.1$ and $\delta_2 = 2.2$. It shows (a,b) the probability to be in a particular state, (c,d) the prevalence $y(t)$ and the average fraction of recovered nodes $r(t)$ for (a,c) infection rate $\beta_{21} = 1$ and (b,d) $\beta_{21} = 4$.}
    \label{fig_SIR_numerics}
\end{figure*}

\subsection{Solution on a complete graph with $N=3$ nodes}
For $N=3$ nodes, the contact graph is a triangle. For simplicity, we assume that the contact graph is symmetric: $\beta_{23} = \beta_{32}$. Its Markov graph contains 27 states and is visualised in Figure~\ref{fig_markov_graph_SIR} in the main text. Starting in the initial state, that is, node 1 being infected and node 2 and 3 being susceptible, the exact solution equals
\begin{align*}
    s_{0}(t) &= 0 \\
    s_{1}(t) &= e^{-(\beta_{21}+\beta_{31}+\delta_1)t} \\
    s_{2}(t) &= \frac{\delta_1}{\beta_{21}+\beta_{31}+\delta_1} \left( 1 - e^{-(\beta_{21}+\beta_{31}+\delta_1) t} \right) \\
    s_{3}(t) &= 0 \\
    s_{4}(t) &= \frac{\beta_{21}}{\beta_{32} - \beta_{21} + \delta_2} \left( e^{-(\beta_{21} + \beta_{31} + \delta_1)t} - e^{-(\beta_{31} + \beta_{32} + \delta_1 + \delta_2)t} \right) \\
    s_5(t) &= \frac{\beta_{21} \delta_1}{(\beta_{31} + \delta_1)(\beta_{21} + \beta_{31} - \beta_{32} + \delta_1 - \delta_2)} e^{-(\beta_{32} + \delta_2)t} \\
    &\qquad +\frac{\beta_{21} \delta_1}{(\beta_{21} - \beta_{32} - \delta_2)(\beta_{21} + \beta_{31} - \beta_{32} + \delta_1 - \delta_2)} e^{-(\beta_{21} + \beta_{31} + \delta_1)t} \\
    &\qquad + \frac{\beta_{21} \delta_1}{(\beta_{31} + \delta_1)( - \beta_{21} + \beta_{32} + \delta_2)} e^{-(\beta_{31} + \beta_{32} + \delta_1 + \delta_2)t} \\
    s_{6}(t) &= 0 \\
    s_{7}(t) &= \frac{\delta_2}{\beta_{32}+\delta_2} e^{-(\beta_{31}+\delta_1)t} + \frac{\delta_2}{\beta_{21}-\beta_{32}-\delta_2} e^{-(\beta_{21}+\beta_{31}+\delta_1)t} + \frac{\beta_{21} \delta_2}{(\beta_{32}+\delta_2)(-\beta_{21} + \beta_{32} + \delta_2)} e^{-(\beta_{31} + \beta_{32} + \delta_1 + \delta_2)t} \\
    s_{8}(t) &= \frac{\beta_{21} \delta_1 \delta_2}{(\beta_{31}+\delta_1) (\beta_{21}+\beta_{31}+\delta_1) (\beta_{32}+\delta_2)} - \frac{\delta_1 \delta_2 }{(\beta_{31}+\delta_1) (\beta_{32}+\delta_2)} e^{-(\beta_{31}+\delta_1) t} \\
    &\qquad-\frac{\beta_{21} \delta_1 \delta_2}{(\beta_{31}+\delta_1) (\beta_{21}+\beta_{31}-\beta_{32}+\delta_1-\delta_2) (\beta_{32}+\delta_2)} e^{-(\beta_{32}+\delta_2) t} \\
    &\qquad-\frac{\delta_1 \delta_2 (2 \beta_{21}+\beta_{31}-\beta_{32}+\delta_1-\delta_2) }{(\beta_{21}+\beta_{31}+\delta_1) (\beta_{21}-\beta_{32}-\delta_2) (\beta_{21}+\beta_{31}-\beta_{32}+\delta_1-\delta_2)} e^{-(\beta_{21}+\beta_{31}+\delta_1) t} \\
    &\qquad+\frac{\beta_{21} \delta_1 \delta_2}{(\beta_{31}+\delta_1) (\beta_{21}-\beta_{32}-\delta_2) (\beta_{32}+\delta_2)} e^{-(\beta_{31}+\beta_{32}+\delta_1+\delta_2) t} \\
    s_{9}(t) &= 0 \\
    s_{10}(t) &= \frac{\beta_{31}}{\beta_{31} - \beta_{32} - \delta_3} \left(- e^{-(\beta_{21} + \beta_{31} + \delta_1)t} + e^{-(\beta_{21} + \beta_{32} + \delta_1 + \delta_2)t} \right) \\
    s_{11}(t) &= \frac{\beta_{31} \delta_1}{(\beta_{21}+\delta_1) (\beta_{21}+\beta_{31}-\beta_{32}+\delta_1-\delta_3)} e^{-(\beta_{32}+\delta_3) t} \\
    &\qquad+\frac{\beta_{31} \delta_1}{(\beta_{31}-\beta_{32}-\delta_3) (\beta_{21}+\beta_{31}-\beta_{32}+\delta_1-\delta_3)} e^{-(\beta_{21}+\beta_{31}+\delta_1) t} \\
    &\qquad- \frac{\beta_{31} \delta_1}{(\beta_{21}+\delta_1) (\beta_{31}-\beta_{32}-\delta_3)} e^{-(\beta_{21}+\beta_{32}+\delta_1+\delta_3) t} \\
    s_{12}(t) &= 0 \\
    s_{13}(t) &= \frac{-\beta_{21}^2 \beta_{31}-\beta_{21} \beta_{31}^2+\beta_{21} \beta_{32}^2+\beta_{31} \beta_{32}^2+\beta_{21} \beta_{31} \delta_2+\beta_{31} \beta_{32} \delta_2+\beta_{21} \beta_{31} \delta_3+\beta_{21} \beta_{32} \delta_3}{(\beta_{21}-\beta_{32}-\delta_2) (\beta_{21}+\beta_{31}-\delta_2-\delta_3)(-\beta_{31}+\beta_{32}+\delta_3)} e^{-(\beta_{21}+\beta_{31}+\delta_1) t} \\
    &\qquad+\frac{\beta_{31} (-\beta_{21}-\beta_{32})}{(\beta_{21}+\beta_{32}-\delta_2) (\beta_{31}-\beta_{32}-\delta_3)} e^{-(\beta_{21}+\beta_{32}+\delta_1+\delta_3)t} \\
    &\qquad+\frac{\beta_{21} (-\beta_{31}-\beta_{32})}{(\beta_{21}-\beta_{32}-\delta_2) (\beta_{31}+\beta_{32}-\delta_3)} e^{-(\beta_{31}+\beta_{32}+\delta_1+\delta_2)t} \\
    &\qquad+\Bigg(\beta_{21}^2 \beta_{31}+\beta_{21} \beta_{31}^2+\beta_{21}^2
    \beta_{32}+2 \beta_{21} \beta_{31} \beta_{32}+\beta_{31}^2 \beta_{32}+\beta_{21} \beta_{32}^2+\beta_{31} \beta_{32}^2-\beta_{21} \beta_{31} \delta_2-\beta_{21}
    \beta_{32} \delta_2 \\
    &\qquad-\beta_{21} \beta_{31} \delta_3-\beta_{31} \beta_{32} \delta_3\Bigg) \Bigg/ \Bigg((\beta_{21}+\beta_{32}-\delta_2)
    (\beta_{31}+\beta_{32}-\delta_3) (\beta_{21}+\beta_{31}-\delta_2-\delta_3) \Bigg) e^{-(\delta_1+\delta_2+\delta_3) t}
\end{align*}
\begin{align*}
    s_{14}(t) &= \Bigg(\beta_{21}^4 \beta_{31} \delta_1+3 \beta_{21}^3 \beta_{31}^2 \delta_1+3 \beta_{21}^2 \beta_{31}^3 \delta_1+\beta_{21}
    \beta_{31}^4 \delta_1-2 \beta_{21}^3 \beta_{32}^2 \delta_1-7 \beta_{21}^2 \beta_{31} \beta_{32}^2 \delta_1-7 \beta_{21} \beta_{31}^2 \beta_{32}^2
    \delta_1 \\
    &\qquad-2 \beta_{31}^3 \beta_{32}^2 \delta_1+3 \beta_{21}^2 \beta_{32}^3 \delta_1+6 \beta_{21} \beta_{31} \beta_{32}^3 \delta_1+3 \beta_{31}^2
    \beta_{32}^3 \delta_1-\beta_{21} \beta_{32}^4 \delta_1-\beta_{31} \beta_{32}^4 \delta_1+2 \beta_{21}^3 \beta_{31} \delta_1^2 \\
    &\qquad+4 \beta_{21}^2 \beta_{31}^2 \delta_1^2+2 \beta_{21} \beta_{31}^3 \delta_1^2-3 \beta_{21}^2 \beta_{32}^2 \delta_1^2-6 \beta_{21} \beta_{31} \beta_{32}^2 \delta_1^2-3 \beta_{31}^2 \beta_{32}^2 \delta_1^2+2 \beta_{21} \beta_{32}^3 \delta_1^2+2 \beta_{31} \beta_{32}^3 \delta_1^2 \\
    &\qquad+\beta_{21}^2 \beta_{31} \delta_1^3+\beta_{21} \beta_{31}^2 \delta_1^3-\beta_{21} \beta_{32}^2 \delta_1^3-\beta_{31} \beta_{32}^2 \delta_1^3-2 \beta_{21}^3 \beta_{31} \delta_1 \delta_2-4
    \beta_{21}^2 \beta_{31}^2 \delta_1 \delta_2-2 \beta_{21} \beta_{31}^3 \delta_1 \delta_2 \\
    &\qquad-2 \beta_{21}^2 \beta_{31} \beta_{32} \delta_1 \delta_2-4
    \beta_{21} \beta_{31}^2 \beta_{32} \delta_1 \delta_2-2 \beta_{31}^3 \beta_{32} \delta_1 \delta_2+2 \beta_{21}^2 \beta_{32}^2 \delta_1 \delta_2+9\beta_{21} \beta_{31} \beta_{32}^2 \delta_1 \delta_2+6 \beta_{31}^2 \beta_{32}^2 \delta_1 \delta_2 \\
    &\qquad-2 \beta_{21} \beta_{32}^3 \delta_1 \delta_2-3\beta_{31} \beta_{32}^3 \delta_1 \delta_2-3 \beta_{21}^2 \beta_{31} \delta_1^2 \delta_2-3 \beta_{21} \beta_{31}^2 \delta_1^2 \delta_2-3 \beta_{21} \beta_{31} \beta_{32} \delta_1^2 \delta_2-3 \beta_{31}^2 \beta_{32} \delta_1^2 \delta_2 \\
    &\qquad+2 \beta_{21} \beta_{32}^2 \delta_1^2 \delta_2+4 \beta_{31} \beta_{32}^2 \delta_1^2 \delta_2-\beta_{21} \beta_{31} \delta_1^3 \delta_2-\beta_{31} \beta_{32} \delta_1^3 \delta_2+\beta_{21}^2 \beta_{31} \delta_1 \delta_2^2+\beta_{21} \beta_{31}^2 \delta_1 \delta_2^2 \\
    &\qquad+3 \beta_{21} \beta_{31} \beta_{32} \delta_1 \delta_2^2+3 \beta_{31}^2 \beta_{32} \delta_1 \delta_2^2-3 \beta_{31} \beta_{32}^2 \delta_1 \delta_2^2+\beta_{21} \beta_{31} \delta_1^2 \delta_2^2+2 \beta_{31} \beta_{32} \delta_1^2 \delta_2^2-\beta_{31} \beta_{32} \delta_1 \delta_2^3 \\
    &\qquad-2 \beta_{21}^3 \beta_{31} \delta_1 \delta_3-4 \beta_{21}^2 \beta_{31}^2 \delta_1 \delta_3-2 \beta_{21} \beta_{31}^3 \delta_1 \delta_3-2 \beta_{21}^3 \beta_{32} \delta_1 \delta_3-4 \beta_{21}^2 \beta_{31} \beta_{32} \delta_1 \delta_3-2 \beta_{21} \beta_{31}^2 \beta_{32} \delta_1 \delta_3 \\
    &\qquad+6 \beta_{21}^2 \beta_{32}^2 \delta_1 \delta_3+9 \beta_{21} \beta_{31} \beta_{32}^2 \delta_1 \delta_3+2 \beta_{31}^2 \beta_{32}^2 \delta_1 \delta_3-3 \beta_{21} \beta_{32}^3 \delta_1 \delta_3-2 \beta_{31} \beta_{32}^3 \delta_1 \delta_3-3\beta_{21}^2 \beta_{31} \delta_1^2 \delta_3 \\
    &\qquad-3 \beta_{21} \beta_{31}^2 \delta_1^2 \delta_3-3 \beta_{21}^2 \beta_{32} \delta_1^2 \delta_3-3 \beta_{21} \beta_{31} \beta_{32} \delta_1^2 \delta_3+4 \beta_{21} \beta_{32}^2 \delta_1^2 \delta_3+2 \beta_{31} \beta_{32}^2 \delta_1^2 \delta_3-\beta_{21} \beta_{31} \delta_1^3 \delta_3 \\
    &\qquad-\beta_{21} \beta_{32} \delta_1^3 \delta_3+3 \beta_{21}^2 \beta_{31} \delta_1 \delta_2 \delta_3+3 \beta_{21} \beta_{31}^2 \delta_1 \delta_2 \delta_3+2 \beta_{21}^2 \beta_{32} \delta_1 \delta_2 \delta_3+4 \beta_{21} \beta_{31} \beta_{32} \delta_1 \delta_2 \delta_3 \\
    &\qquad+2 \beta_{31}^2 \beta_{32} \delta_1 \delta_2 \delta_3-4 \beta_{21} \beta_{32}^2 \delta_1 \delta_2 \delta_3-4 \beta_{31} \beta_{32}^2 \delta_1 \delta_2 \delta_3+2 \beta_{21} \beta_{31} \delta_1^2 \delta_2 \delta_3+2 \beta_{21} \beta_{32} \delta_1^2 \delta_2 \delta_3 \\
    &\qquad+2 \beta_{31} \beta_{32} \delta_1^2 \delta_2 \delta_3-\beta_{21} \beta_{31} \delta_1 \delta_2^2 \delta_3-2 \beta_{31} \beta_{32} \delta_1 \delta_2^2 \delta_3+\beta_{21}^2 \beta_{31} \delta_1 \delta_3^2+\beta_{21} \beta_{31}^2 \delta_1 \delta_3^2+3 \beta_{21}^2 \beta_{32} \delta_1 \delta_3^2 \\
    &\qquad+3 \beta_{21} \beta_{31} \beta_{32} \delta_1 \delta_3^2-3 \beta_{21} \beta_{32}^2 \delta_1 \delta_3^2+\beta_{21} \beta_{31} \delta_1^2 \delta_3^2+2 \beta_{21} \beta_{32} \delta_1^2 \delta_3^2-\beta_{21} \beta_{31} \delta_1 \delta_2 \delta_3^2-2 \beta_{21} \beta_{32} \delta_1 \delta_2 \delta_3^2 \\
    &\qquad-\beta_{21} \beta_{32} \delta_1 \delta_3^3\Bigg) \Bigg/\Bigg((\beta_{21}+\beta_{31}-\beta_{32}+\delta_1-\delta_2)
    (-\beta_{21}+\beta_{32}+\delta_2) (\beta_{31}-\beta_{32}-\delta_3) \\
    &\qquad \times (\beta_{21}+\beta_{31}-\beta_{32}+\delta_1-\delta_3) (\beta_{21}+\beta_{31}-\delta_2-\delta_3)
    (\beta_{21}+\beta_{31}+\delta_1-\delta_2-\delta_3) \Bigg) e^{-(\beta_{21}+\beta_{31}+\delta_1) t} \\
    &\qquad -\frac{\beta_{21} \beta_{32} \delta_1}{(\beta_{31}+\delta_1)(\beta_{21}+\beta_{31}-\beta_{32}+\delta_1-\delta_2) (\beta_{32}-\delta_3)} e^{-(\beta_{32}+\delta_2) t} \\
    &\qquad-\frac{\beta_{31} \beta_{32} \delta_1}{(\beta_{21}+\delta_1) (\beta_{32}-\delta_2)(\beta_{21}+\beta_{31}-\beta_{32}+\delta_1-\delta_3)} e^{-(\beta_{32}+\delta_3) t} \\
    &\qquad+\frac{\beta_{31} \left(\beta_{21}^2 \delta_1+2 \beta_{21} \beta_{32} \delta_1+\beta_{32}^2 \delta_1+\beta_{21} \delta_1^2+\beta_{32} \delta_1^2-\beta_{32} \delta_1 \delta_2\right)}{(\beta_{21}+\delta_1) (\beta_{21}+\beta_{32}-\delta_2) (\beta_{21}+\beta_{32}+\delta_1-\delta_2)(\beta_{31}-\beta_{32}-\delta_3)} e^{-(\beta_{21}+\beta_{32}+\delta_1+\delta_3)t} \\
    &\qquad+\frac{\beta_{21} \left(\beta_{31}^2 \delta_1+2 \beta_{31} \beta_{32} \delta_1+\beta_{32}^2 \delta_1+\beta_{31} \delta_1^2+\beta_{32} \delta_1^2-\beta_{32} \delta_1 \delta_3\right)}{(\beta_{31}+\delta_1) (\beta_{21}-\beta_{32}-\delta_2) (\beta_{31}+\beta_{32}-\delta_3) (\beta_{31}+\beta_{32}+\delta_1-\delta_3)} e^{-(\beta_{31}+\beta_{32}+\delta_1+\delta_2) t} \\
    &\qquad+\Bigg(\frac{\beta_{31} (\beta_{21}+\beta_{32})}{(\beta_{21}+\beta_{32}+\delta_1-\delta_2) (\beta_{21}+\beta_{31}+\delta_1-\delta_2-\delta_3)} \\
    &\qquad+\frac{\beta_{31}\beta_{32} \delta_1}{(\beta_{32}-\delta_2) (\beta_{21}+\beta_{32}+\delta_1-\delta_2) (\beta_{21}+\beta_{31}+\delta_1-\delta_2-\delta_3)} \\
    &\qquad+\frac{\beta_{21}(\beta_{31}+\beta_{32})}{(\beta_{31}+\beta_{32}+\delta_1-\delta_3) (\beta_{21}+\beta_{31}+\delta_1-\delta_2-\delta_3)} \\
    &\qquad-\frac{\beta_{21} \beta_{32}\delta_1}{(\beta_{32}-\delta_3) (\beta_{31}+\beta_{32}+\delta_1-\delta_3) (-\beta_{21}-\beta_{31}-\delta_1+\delta_2+\delta_3)}\Bigg) e^{-(\delta_2+\delta_3)t} \\
    &\qquad- \Bigg(\beta_{21}^2 \beta_{31}+\beta_{21} \beta_{31}^2+\beta_{21}^2 \beta_{32}+2 \beta_{21} \beta_{31} \beta_{32}+\beta_{31}^2 \beta_{32}+\beta_{21} \beta_{32}^2+\beta_{31} \beta_{32}^2-\beta_{21} \beta_{31} \delta_2 -\beta_{21} \beta_{32} \delta_2 \\
    &\qquad-\beta_{21} \beta_{31} \delta_3-\beta_{31} \beta_{32} \delta_3\Bigg) \Bigg/ \Bigg((\beta_{21}+\beta_{32}-\delta_2) (\beta_{31}+\beta_{32}-\delta_3) (\beta_{21}+\beta_{31}-\delta_2-\delta_3) \Bigg) e^{-(\delta_1+\delta_2+\delta_3) t} \\
%		s_{15}(t) &= 0 \\
%		s_{16}(t) &=  \\
%		s_{17}(t) &=  \\
%		s_{18}(t) &= 0 \\
%		s_{19}(t) &=  \\
%		s_{20}(t) &=  \\
%		s_{21}(t) &= 0 \\
%		s_{22}(t) &=  \\
%		s_{23}(t) &=  \\
%		s_{24}(t) &= 0 \\
%		s_{25}(t) &=  \\
%		s_{26}(t) &= \frac{}{} e^{-()t}
\end{align*}
We have omitted the remaining states $s_{15}, \ldots, s_{26}$, since they are rather involved. Instead, the formula for the prevalence $y(t)$ is shown below, which is, unfortunately, also rather involved.
\begin{align*}
    3 y(t) &= e^{-\delta_1 t}-\frac{\beta_{21} \delta_3}{(\beta_{21}+\delta_1-\delta_2) (\beta_{32}+\delta_3)} e^{-(\beta_{21}+\delta_1) t} - \frac{\beta_{31} \delta_2}{(\beta_{32}+\delta_2) (\beta_{31}+\delta_1-\delta_3)} e^{-(\beta_{31}+\delta_1) t} \\
    &\qquad+\Bigg(\beta_{32} (-\beta_{21}^5+\beta_{21}^4
    (-5 \beta_{31}+4 \beta_{32}-3 \delta_1+2 \delta_2+3 \delta_3)-\beta_{31} (\beta_{31}-\beta_{32}+\delta_1-\delta_2) \\
    &\qquad\times (\beta_{31}-\beta_{32}-\delta_3)
    \left(\beta_{31}^2+(\delta_1-2 \delta_2) (\delta_1-\delta_3)+\beta_{32} (-2 \delta_1+\delta_2+\delta_3)-\beta_{31} (2 \beta_{32}-2 \delta_1+2
    \delta_2+\delta_3)\right) \\
    &\qquad-\beta_{21}^3 (10 \beta_{31}^2+5 \beta_{32}^2+3 \delta_1^2-5 \delta_1 \delta_2+\delta_2^2-6 \delta_1 \delta_3+6
    \delta_2 \delta_3+2 \delta_3^2+\beta_{32} (-9 \delta_1+6 \delta_2+8 \delta_3) \\
    &\qquad-\beta_{31} (16 \beta_{32}-10 \delta_1+9 \delta_2+11 \delta_3))+\beta_{21}^2 (-10 \beta_{31}^3+2 \beta_{32}^3-\delta_1^3+4 \delta_1^2 \delta_2-2 \delta_1 \delta_2^2+3 \delta_1^2 \delta_3 \\
    &\qquad-9 \delta_1 \delta_2 \delta_3+3
    \delta_2^2 \delta_3-2 \delta_1 \delta_3^2+4 \delta_2 \delta_3^2+\beta_{32}^2 (-8 \delta_1+5 \delta_2+6 \delta_3)+\beta_{32} (6 \delta_1^2-10
    \delta_1 \delta_2+2 \delta_2^2\\
    &\qquad-11 \delta_1 \delta_3+11 \delta_2 \delta_3+3 \delta_3^2)+\beta_{31}^2 (24 \beta_{32}-14 \delta_1+15
    (\delta_2+\delta_3))-\beta_{31} (15 \beta_{32}^2+5 \delta_1^2-14 \delta_1 \delta_2 \\
    &\qquad+4 \delta_2^2-15 \delta_1 \delta_3+18 \delta_2 \delta_3+5
    \delta_3^2+\beta_{32} (-25 \delta_1+20 \delta_2+22 \delta_3)))-\beta_{21} (5 \beta_{31}^4-\beta_{31}^3 (16 \beta_{32}-10 \delta_1 \\
    &\qquad+11 \delta_2+9 \delta_3)+(\beta_{32}+\delta_2) (\beta_{32}-\delta_1+\delta_3) ((\delta_1-\delta_2) (\delta_1-2 \delta_3)+\beta_{32} (-2 \delta_1+\delta_2+\delta_3)) \\
    &\qquad+\beta_{31}^2
    (15 \beta_{32}^2+5 \delta_1^2-15 \delta_1 \delta_2+5 \delta_2^2-14 \delta_1 \delta_3+18 \delta_2 \delta_3+4 \delta_3^2+\beta_{32} (-25
    \delta_1+22 \delta_2+20 \delta_3)) \\
    &\qquad-\beta_{31} (4 \beta_{32}^3+5 \delta_1^2 (\delta_2+\delta_3)+7 \delta_2 \delta_3 (\delta_2+\delta_3)-4 \delta_1 \left(\delta_2^2+4 \delta_2 \delta_3+\delta_3^2\right)+\beta_{32}^2 (-16 \delta_1+11 (\delta_2+\delta_3)) \\
    &\qquad+\beta_{32} (10 \delta_1^2+5 \delta_2^2+22 \delta_2 \delta_3+5 \delta_3^2-20 \delta_1 (\delta_2+\delta_3))))) \Bigg)\Bigg/\Bigg((\beta_{21}-\beta_{32}-\delta_2) (\beta_{21}+\beta_{31}+\delta_1-\delta_2) \\
    &\qquad\times (\beta_{21}+\beta_{31}-\beta_{32}+\delta_1-\delta_2)
    (\beta_{21}+\beta_{31}+\delta_1-\delta_3) (\beta_{21}+\beta_{31}-\beta_{32}+\delta_1-\delta_3) \\
    &\qquad\times (-\beta_{31}+\beta_{32}+\delta_3)\Bigg) e^{-(\beta_{21}+\beta_{31}+\delta_1)t} \\
    &\qquad+\Bigg((-\beta_{21}^3
    (\beta_{32}-\delta_2+\delta_3)-\beta_{21}^2 (\beta_{31}+\beta_{32}+2 \delta_1-2 \delta_2+\delta_3) (\beta_{32}-\delta_2+\delta_3) \\
    &\qquad-\beta_{31} \beta_{32} (\delta_1-\delta_2) (\beta_{32}+\delta_1-\delta_2+\delta_3)-\beta_{21} ((\delta_1-\delta_2) (\beta_{32}-\delta_2+\delta_3)
    (\beta_{32}+\delta_1-\delta_2+\delta_3) \\
    &\qquad+\beta_{31} (\beta_{32}^2+2 \beta_{32} (\delta_1-\delta_2+\delta_3)-(\delta_2-\delta_3) (\delta_1-\delta_2+\delta_3))))\Bigg)\Bigg/\Bigg((\beta_{21}+\delta_1-\delta_2) \\
    &\qquad\times (\beta_{21}+\beta_{31}+\delta_1-\delta_2) (-\beta_{32}+\delta_2-\delta_3) (\beta_{21}+\beta_{32}+\delta_1-\delta_2+\delta_3)\Bigg) e^{-\delta_2 t} \\
    &\qquad+\Bigg((-\beta_{21} (\beta_{31} (\beta_{32}^2+2
    \beta_{32} (\delta_1+\delta_2-\delta_3)+(\delta_2-\delta_3) (\delta_1+\delta_2-\delta_3))+\beta_{31}^2 (\beta_{32}+\delta_2-\delta_3) \\
    &\qquad+\beta_{32}(\delta_1-\delta_3)(\beta_{32}+\delta_1+\delta_2-\delta_3))-\beta_{31} (\beta_{31}+\delta_1-\delta_3) (\beta_{32}+\delta_2-\delta_3)
    (\beta_{31}+\beta_{32}+\delta_1+\delta_2-\delta_3))\Bigg)\Bigg/ \\
    &\qquad\Bigg((\beta_{31}+\delta_1-\delta_3) (\beta_{21}+\beta_{31}+\delta_1-\delta_3)
    (\beta_{31}+\beta_{32}+\delta_1+\delta_2-\delta_3) (-\beta_{32}-\delta_2+\delta_3)\Bigg) e^{-\delta_3 t} \\
    &\qquad-\frac{\beta_{21} \beta_{32} \delta_1}{(\beta_{31}+\delta_1)(\beta_{21}+\beta_{31}-\beta_{32}+\delta_1-\delta_2) (\beta_{32}+\delta_2-\delta_3)} e^{-(\beta_{32}+\delta_2) t} \\
    &\qquad-\frac{\beta_{31} \beta_{32} \delta_1}{(\beta_{21}+\delta_1)(\beta_{21}+\beta_{31}-\beta_{32}+\delta_1-\delta_3) (\beta_{32}-\delta_2+\delta_3)} e^{-(\beta_{32}+\delta_3)t} \\
    &\qquad-\frac{\beta_{21} \beta_{31} \beta_{32} (\beta_{21}+\beta_{32}+\delta_1+\delta_3)}{(\beta_{21}+\delta_1) (\beta_{31}-\beta_{32}-\delta_3) (\beta_{32}+\delta_3) (\beta_{21}+\beta_{32}+\delta_1-\delta_2+\delta_3)} e^{-(\beta_{21}+\beta_{32}+\delta_1+\delta_3) t} \\
    &\qquad-\frac{\beta_{21} \beta_{31} \beta_{32} (\beta_{31}+\beta_{32}+\delta_1+\delta_2)}{(\beta_{31}+\delta_1) (\beta_{21}-\beta_{32}-\delta_2) (\beta_{32}+\delta_2) (\beta_{31}+\beta_{32}+\delta_1+\delta_2-\delta_3)} e^{-(\beta_{31}+\beta_{32}+\delta_1+\delta_2) t} \\
\end{align*}

\subsection{Solution on a path graph with $N=3$ nodes}
If the contact graph equals a path graph on $N=3$ nodes, the equation for the prevalence $y(t)$ simplifies to
\begin{align*}
    y(t) &= \frac{1}{3} \Bigg( e^{-\delta_1 t} + \frac{\beta_{21}}{\beta_{21}+\delta_1-\delta_2} e^{-\delta_2 t} + \frac{\beta_{21} \beta_{32}}{(\beta_{21}+\delta_1-\delta_3)(\beta_{32}+\delta_2-\delta_3)} e^{-\delta_3 t} \\
    &\qquad - \frac{\beta_{21} \left( \beta_{21}^2 + (\delta_1-\delta_2)(\delta_1-\delta_3) + \beta_{32} (-2\delta_1+\delta_2+\delta_3) - \beta_{21}(2\beta_{32}-2\delta_1+\delta_2+\delta_3)\right)}{(\beta_{21}+\delta_1-\delta_2)(\beta_{21}-\beta_{32}+\delta_1-\delta_2)(\beta_{21}+\delta_1-\delta_3)} e^{-(\beta_{21}+\delta_1)t} \\
    &\qquad - \frac{\beta_{21} \beta_{32}}{(\beta_{21}-\beta_{32}+\delta_1-\delta_2)(\beta_{32}+\delta_2-\delta_3)} e^{-(\beta_{32}+\delta_2)t} \Bigg)
\end{align*}
and the average fraction of recovered nodes follows as
\begin{align*}
    r(t) &= \frac{1}{3} \Bigg( 1 + \frac{\beta_{21}(2\beta_{32}+\delta_2)}{(\beta_{21}+\delta_1)(\beta_{32}+\delta_2)} - e^{-\delta_1 t} -\frac{\beta_{21}}{\beta_{21}+\delta_1-\delta_2} e^{-\delta_2 t} -\frac{\beta_{21} \beta_{32}}{(\beta_{21}+\delta_1-\delta_3) (\beta_{32}+\delta_2-\delta_3)} e^{-\delta_3 t} \\
    &\qquad+ \Bigg(\beta_{21} (\beta_{21}^2 \delta_2+(\delta_1-\delta_2) \delta_2 (\delta_1-\delta_3)+2 \beta_{32} \delta_2 \delta_3-\beta_{32} \delta_1 (\delta_2+\delta_3)-\beta_{21} (\beta_{32} (\delta_2+\delta_3) \\
    &\qquad+\delta_2 (-2 \delta_1+\delta_2+\delta_3)))\Bigg) \Bigg/ \Bigg((\beta_{21}+\delta_1) (\beta_{21}+\delta_1-\delta_2) (\beta_{21}-\beta_{32}+\delta_1-\delta_2) (\beta_{21}+\delta_1-\delta_3) \Bigg) e^{-(\beta_{21}+\delta_1)t} \\
    &\qquad + \frac{\beta_{21} \beta_{32} \delta_3}{(\beta_{21}-\beta_{32}+\delta_1-\delta_2) (\beta_{32}+\delta_2) (\beta_{32}+\delta_2-\delta_3)} e^{-(\beta_{32}+\delta_2)t} \Bigg)
\end{align*}

\end{widetext}

\end{document}